\documentclass[11pt,draftclsnofoot,onecolumn,twoside]{IEEEtran}

\usepackage{amsmath}
\usepackage{amssymb}
\usepackage{latexsym}
\usepackage{verbatim}
\usepackage[final]{graphicx}
\usepackage{psfrag}

\newtheorem{theorem}{Theorem}
\newtheorem{proposition}{Proposition}
\newtheorem{lemma}{Lemma}

\newtheorem{definition}{Definition}

\newenvironment{remark}{\textit{Remark:}}

{\begin{list}               
    {$\bullet$ \hfill}{
        \setlength{\leftmargin}{\parindent}
        \setlength{\parsep}{0.04\baselineskip}
        \setlength{\itemsep}{0.5\parsep}
        \setlength{\labelwidth}{\leftmargin}
        \setlength{\labelsep}{0em}}
    }
{\end{list}}

\providecommand{\eref}[1]{\eqref{eq:#1}}  
\providecommand{\cref}[1]{Chapter~\ref{chap:#1}}
\providecommand{\sref}[1]{Section~\ref{sec:#1}}
\providecommand{\fref}[1]{Figure~\ref{fig:#1}}

\providecommand{\RR}{\ensuremath{\mathbb{R}}}

\providecommand{\HH}{\ensuremath{\mathcal{H}}}
\providecommand{\GG}{\ensuremath{\mathcal{G}}}

\providecommand{\abs}[1]{\lvert#1\rvert}
\providecommand{\norm}[1]{\lVert#1\rVert}

\providecommand{\set}[1]{\left\{#1\right\}}

\providecommand{\bydef}{\overset{\text{def}}{=}}

\renewcommand{\vec}[1]{\ensuremath{\boldsymbol{#1}}}
\providecommand{\mat}[1]{\ensuremath{\boldsymbol{#1}}}

\providecommand{\calA}{\mathcal{A}}
\providecommand{\calB}{\mathcal{B}}
\providecommand{\calS}{\mathcal{S}}

\providecommand{\calN}{\mathcal{N}}

\providecommand{\mA}{\mat{A}} \providecommand{\mB}{\mat{B}}
 \providecommand{\mH}{\mat{H}}
\providecommand{\mD}{\mat{D}} \providecommand{\mF}{\mat{F}}
\providecommand{\mI}{\mat{I}}  
 \providecommand{\mL}{\mat{L}} 
\providecommand{\mM}{\mat{M}} \providecommand{\mP}{\mat{P}} 
 
\providecommand{\mS}{\mat{S}}  
 
\providecommand{\mX}{\mat{X}}
\providecommand{\mZ}{\mat{Z}}

\providecommand{\mLambda}{\mat{\Lambda}}

\providecommand{\vf}{\vec{f}}
\providecommand{\vg}{\vec{g}}

 \providecommand{\vw}{\vec{w}}
\providecommand{\vx}{\vec{x}} \providecommand{\vy}{\vec{y}}
\providecommand{\vz}{\vec{z}} 
 
\providecommand{\vone}{\vec{1}}
 \providecommand{\vv}{\vec{v}}

\providecommand{\op}[1]{\operatorname{#1}}

\newcommand{\LL}{\mathcal{L}}

\newcommand{\DS}{\Delta_s^2}
\newcommand{\DG}{\Delta_{g,u_0}^2}
\newcommand{\DD}{\mathcal{D}_{u_0}}
\providecommand{\gu}{\ensuremath{\gamma_{u_0}}}
\providecommand{\tgu}{\ensuremath{\tilde{\gamma}_{u_0}}}
\providecommand{\etau}{\ensuremath{\eta_{u_0}}}
\providecommand{\deg}{\ensuremath{\operatorname{deg}}}
\providecommand{\vimp}[1]{\ensuremath{\vec{\delta}_{#1}}}

\providecommand{\Erdos}{Erd\H{o}s}
\providecommand{\Renyi}{R\'{e}nyi}

\usepackage{amsmath}
\usepackage{amsfonts}
\usepackage{graphicx}
\usepackage{psfrag}
\usepackage{booktabs}
\usepackage{multirow}
\usepackage{color}
\usepackage[nosort]{cite}
\usepackage{afterpage}
\usepackage{rotating}
\usepackage{algorithmic}
\usepackage{algorithm}
\usepackage{float}
\usepackage{mathtools}
\usepackage{enumerate}
\usepackage{lineno}
\usepackage{url}

\usepackage{subfigure}

\DeclareMathOperator{\diag}{diag}

\floatname{algorithm}{Procedure}

\newcommand{\Ss}{\mathcal{S}}

\begin{document}

\title{A Spectral Graph Uncertainty Principle}

\author{Ameya~Agaskar%
            ,~\IEEEmembership{Student Member,~IEEE,} 
        and~Yue~M.~Lu,~\IEEEmembership{Senior Member,~IEEE}%
\thanks{The authors are with the Signals, Information, and Networks Group (SING) at the School of Engineering and Applied Sciences, 
          Harvard University, Cambridge, MA 02138, USA 
        (e-mail: \{aagaskar, yuelu\}@seas.harvard.edu).}
\thanks{A.~Agaskar is also with MIT Lincoln Laboratory. The Lincoln Laboratory portion of this work is sponsored by the Department of the Air Force under
            Contract FA8721-05-C-0002. Opinions, interpretations, conclusions and
            recommendations are those of the authors and are not necessarily
            endorsed by the United States Government.}
\thanks{The material in this paper was presented in part at the 2011 SPIE Conference on Wavelets and Sparsity (XIV)
        and the 2012 IEEE International Conference on Acoustics, Speech and Signal Processing (ICASSP).} 
}

\markboth{}{Agaskar and Lu: A Spectral Graph Uncertainty Principle}

\maketitle
\begin{abstract}
The spectral theory of graphs provides a bridge between classical signal processing and the nascent field of graph signal processing. In this paper, a 
spectral graph analogy to Heisenberg's celebrated uncertainty principle is developed. Just as the classical result provides a tradeoff between signal 
localization in time and frequency, this result provides a fundamental tradeoff between a signal's localization on a graph and in its spectral domain.
Using the eigenvectors of the graph Laplacian as a surrogate Fourier basis, quantitative definitions of graph and spectral ``spreads'' are given,
and a complete characterization of the feasibility region of these two quantities is developed.
In particular, the lower boundary of the region, referred to as the uncertainty curve,
is shown to be achieved by eigenvectors associated with the smallest eigenvalues of an affine family of matrices. The convexity of the uncertainty curve allows it 
to be found to within $\varepsilon$ by a fast approximation algorithm requiring $\mathcal{O}(\varepsilon^{-1/2})$ typically sparse eigenvalue evaluations.
Closed-form expressions for the uncertainty curves for some special classes of graphs are derived, 
and an accurate analytical approximation for the expected uncertainty curve of \Erdos-\Renyi{} random graphs is developed. 
These theoretical results are validated by numerical experiments, which also reveal an intriguing connection between 
diffusion processes on graphs and the uncertainty bounds.
\end{abstract}


\ifCLASSOPTIONdraftcls
    \pagebreak[4]
\fi

\section{Introduction}
\label{sec:introduction}

Heisenberg's uncertainty principle is a cornerstone of signal processing. The simple inequality \cite{folland_uncertainty_1997, VK_book},
\begin{equation}
\Delta_t^2 \Delta_\omega^2 \geq \frac{1}{4},
\label{eq:classical_uncertainty}
\end{equation}
in which $\Delta_t^2$ and $\Delta_\omega^2$ measure the ``time spread'' and ``frequency spread'' of some signal, respectively, is one
way to precisely characterize a general principle with far-reaching consequences: that a signal cannot be concentrated in both time and frequency.

In this paper, we establish analogous uncertainty principles for signals defined on graphs. The study of signals on graphs,
and the extension of classical signal processing techniques to such nonstandard domains, has received growing
interest in the past few years
(see, \emph{e.g.}, \cite{buades_non-local_2005, coifman_diffusion_2006, coifman_diffusion_2006b,
hammond_wavelets_2010, khan_higher_2010, cattivelli_distributed_2011, pesenson_sampling_2008,narang_perfect_2012, dimakis_gossip_2010}). 
These studies are often motivated (and enabled) by the deluge of modern data collected on various technological, social, biological, and informational networks \cite{Kolaczyk:2009}. The efficient acquisition, representation, and analysis of such high-dimensional graph-based data present challenges that should be addressed by the development of new signal processing theories and tools.


\subsection{Related Work}

Uncertainty principles date
back to Heisenberg \cite{folland_uncertainty_1997}, who in 1927 proved a result 
that Weyl and Pauli soon afterward generalized to (\ref{eq:classical_uncertainty}). It was also shown that the bound in (\ref{eq:classical_uncertainty}) is achievable by Gaussian-shaped functions
and frequency modulations thereof. A lifetime later, analogous results were found for discrete-time signals as well \cite{ishii_uncertainty_1986,calvez_uncertainty_1992}. Similar uncertainty principles have also been established on the unit sphere $S^d$ \cite{Goh:2004} and, in more
abstract settings, on compact Riemannian manifolds \cite{Wolfgang:2010}.   

In a different line of work, Donoho and Stark \cite{donoho_uncertainty_1989} introduced a new concept of uncertainty related to signal support size.
They showed that a length $N$ discrete-time signal with support set $\mathcal{T}$ in the time domain and support set  $\mathcal{W}$  in the frequency
domain satisfies $\abs{\mathcal{T}} \, \abs{\mathcal{W}} \geq N$. 
This bound is a \emph{nonlocal} uncertainty principle---it 
limits the cardinality of a signal's time and frequency support sets, even if each is the disjoint union of far-flung subsets.
Further studied in, \emph{e.g}, \cite{donoho_uncertainty_2001,candes_quantitative_2005,candes_robust_2006}, these nonlocal uncertainty principles laid the foundation for sparse signal recovery from partial measurements.

In the same vein of the classical (and local) uncertainty principle stated in \eref{classical_uncertainty}, we have been studying the following question: given an arbitrary graph, to what extent can a signal be simultaneously localized on that graph and in the ``frequency'' domain? To obtain the spectral representation of these signals, we use the standard approach of treating the eigenvectors of the graph 
Laplacian operator \cite{GodsilR:2001} as a Fourier basis. The Laplacian encodes a notion of smoothness on a 
graph \cite{belkin_regularization_2004} and is analogous to the Laplace-Beltrami operator on a manifold \cite{belkin_problems_2003}.

The analogy between the spectral decomposition of graph Laplacians and the standard Fourier transform has been used 
to extend the concept of bandlimited sampling 
to signals defined on graphs \cite{pesenson_sampling_2008}
and in the construction of wavelet transforms on graphs \cite{coifman_diffusion_2006, hammond_wavelets_2010, narang_perfect_2012}.
In the latter case, as pointed out in \cite{narang_perfect_2012}, a desirable property of the wavelet transforms is that the dictionary elements (\emph{i.e.}, wavelets) should be well-localized in the graph and spectral domains. Our results provide a way to precisely quantify this desideratum, as well as its fundamental limit.

\subsection{Contributions and Paper Organization}
\label{subsec:contributions}
We begin in \sref{Math} with a review of some basic concepts in graph theory, including the definition of the graph Laplacian matrix and its spectral decomposition.
After justifying the use of the Laplacian eigenvectors as a Fourier basis on graphs, we define in \sref{spreads} the \emph{graph spread} about a vertex $u_0$, $\DG(\vx)$,
and the \emph{spectral spread}, $\DS(\vx)$, of a signal $\vx$ defined on a graph. These two quantities, which we first introduced in some
preliminary work \cite{agaskar_uncertainty_2011, agaskar_uncertainty_2012}, are defined in analogy to the standard time and frequency spreads, respectively.

In \cite{agaskar_uncertainty_2011}, we developed a lower bound on the product of $\DG$ and $\DS$ analogous to (\ref{eq:classical_uncertainty}). However,
the bound was not tight and applied only under restrictive conditions for the graph and the signal on it. In \cite{agaskar_uncertainty_2012} we took a new approach to
characterize a more general and precise relationship between the two kinds of uncertainty. In this paper, we continue this line of investigation, and provide
a rigorous basis for the arguments presented in \cite{agaskar_uncertainty_2012}, in addition to some new results.

The main contributions of this paper are developed in Section \ref{sec:Bound}, where 
we characterize the uncertainty bound, in Section \ref{sec:special}, where we analyze the bound when applied to special
families of graphs, and in Section \ref{sec:diffusion}, where we reveal a connection between diffusion processes and the uncertainty bound. 
The main results are summarized as follows: \label{page:newcontributions}

    1. \emph{Convexity of the feasible region}:
        We prove that, when the underlying graph is connected and contains at least three
        vertices, the feasibility region of all possible pairs $(\DS(\vx), \DG(\vx))$ is
        a \emph{bounded} and \emph{convex} set. The feasibility region's convexity was stated without proof in \cite{agaskar_uncertainty_2012}.

    2. \emph{Characterization of the uncertainty curve}:
        We provide a complete characterization of the curve
        \[
        \gu(s) = \min_{\vx} \DG(\vx) \text{ subject to } \DS(\vx) = s,
        \]
        which forms the lower boundary of the feasibility region. Studying $\gu(s)$, which we will refer to
        as the \emph{uncertainty curve}, is important because it is a fundamental bound analogous to the classical
        uncertainty bound (\ref{eq:classical_uncertainty}).
        Theorem~\ref{thm:smallesteig} states that each point on the uncertainty 
        curve is achieved by an eigenvector associated with the smallest eigenvalue of a particular 
        matrix-valued function $\mM(\alpha)$. Varying the parameter $\alpha$ allows one to ``trace''
        and obtain the entire curve $\gu(s)$. A rigorous and complete proof of Theorem~\ref{thm:smallesteig} is provided
        in this paper, extending the rough argument given in \cite{agaskar_uncertainty_2012}.
        Based the
        convexity of $\gu(s)$, we show in \sref{curve_approx} that the \emph{sandwich algorithm} \cite{rote_convergence_1992}
        can be used to efficiently produce a piecewise linear approximation 
        for the uncertainty curve that differs from the true curve by at most $\varepsilon$ (under a suitable
        error metric) and 
        requires solving $\mathcal{O}(\varepsilon^{-1/2})$ typically sparse eigenvalue problems.

     3. \emph{Special graph families}: The uncertainty curves for several special families of graphs are investigated in \sref{special}. 
For complete graphs and star graphs, we derive closed-form formulas for the uncertainty curves $\gu(s)$. For \Erdos{}-\Renyi{} random graphs
\cite{Erdos:1959ai,Erdos:1960dq}, we develop an analytical approximation for the expected value of $\gu(s)$, which is shown through experiment
to be very accurate.

4. \emph{Diffusion process on a graph}: In \sref{diffusion}, we reveal an intriguing connection between the classical uncertainty principle for
functions on the real line and our results for signals on graphs. In the classical case, the solution to the heat equation $\frac{du}{dt} = \frac{d^2u}{dy^2}$
    starting at $t=0$ as an impulse
    is a Gaussian function with a variance that grows linearly with $t$; this solution achieves the Heisenberg uncertainty bound (\ref{eq:classical_uncertainty}). 
     We first show experimental results indicating that a diffusion process starting with an impulse on a graph
    follows the graph uncertainty curve very closely (though not, in general, exactly.) We then prove in Proposition \ref{prop:diffusion} that 
    the match is exact for the special cases of a star graph or a complete graph. We further prove in Proposition \ref{prop:diffusion2}
    that for general graphs, under a simple condition on the distance function on the graph,
    the first two derivatives of the uncertainty curve and the curve traced by the diffusion
    process match at the point corresponding to $t=0$.
    We conclude the paper in \sref{conclusions}.

\section{Mathematical Formulation}
\label{sec:Math}
\subsection{Graphs, Signals, and Notation}
\label{subsec:graphdefs}

We define a simple, undirected graph as $G = (V, E)$, where $V = \{v_1, v_2, \ldots, v_N\}$ is a set of $N$ vertices and $E = \{e_1, e_2, \ldots, e_M\}$ is the set of $M$ edges. Each edge is an unordered pair of two different vertices $u, v \in V$, and we will use the notation $u \sim v$ to  indicate that $u$ and $v$ are connected by an edge. The fundamental structure of a graph $G$ can be captured by its \emph{adjacency matrix} $\mA = [a_{ij}]_{ij}$, where $a_{ij} = 1$ if there is an edge between
$v_i$ and $v_j$, and $a_{ij} = 0$ otherwise. As defined, the diagonal of $\mA$ is always zero because a simple graph may contain no loops (\emph{i.e.}, edges
connecting one vertex to itself), and $\mA$ is symmetric because the graph is undirected. (A common generalization is to consider a weighted graph,
where each edge $e_m$ ($1 \le m \le M$) is associated with a positive ``weight'' $w_m$. However, in this paper we only consider unweighted graphs.)

The degree of a vertex $v$, denoted by $\deg(v)$, is the number of edges incident upon that vertex. We define $\mD$ as the diagonal matrix that
has the vertex degrees on its diagonal, \emph{i.e.},
\begin{equation}\label{eq:degree_mtx}
\mD \bydef \text{diag}\set{\deg(v_1), \deg(v_2), \ldots, \deg(v_N)}.
\end{equation}

To quantify the graph-domain spread of a signal, we will need a notion of distance, denoted by $d(u, v)$, between any pair of vertices $u$ and $v$ on the graph. A simple choice is to use the \emph{geodesic distance} \cite{GodsilR:2001}, in which case $d(u, v)$ is the length of the shortest path connecting the two vertices. In this work, we only consider connected graphs, so $d(u, v)$ is always finite. Other distance functions have been proposed in the literature, 
including the resistance distance \cite{klein_resistance_1993} and the diffusion distance \cite{coifman_diffusion_2006}. 
Our subsequent discussions are not confined to any particular choice of the distance function.
The only requirement is that 
$d(u, v)$ should form a semi-metric: namely, $d(u, v) \ge 0$ with equality if and only if $u = v$, and $d(u,v) = d(v,u)$.

A finite-energy signal defined on the graph $\vx \in \ell^2(G)$ is a mapping from the set of vertices to $\RR$.
It can be treated as a vector in $\mathbb{R}^N$, and so any such signal will be denoted by a boldface variable.
There is a natural inner product on $\ell^2(G)$ defined by $\left\langle \vx, \vy \right\rangle = \vy^T \vx$,
which induces a norm $\norm{\vx} = \sqrt{\vx^T \vx}$. We will denote the value of $\vx$ at vertex $v$ by $x(v)$.
An impulse at $v \in V$, \textit{i.e.}, a signal that has value $1$ at $v$ and $0$ everywhere else, will be
denoted as $\vimp{v}$.

\subsection{The Laplacian Matrix and Graph Fourier Transforms}
\label{sec:graphLap}

As mentioned in \sref{introduction}, the graph Laplacian matrix plays an important role in this work. There are several different definitions of the Laplacian matrix commonly used in the literature. The unnormalized Laplacian matrix \cite{GodsilR:2001} is given by $\mL_{\text{unnorm}} \bydef \mD - \mA$, where $\mD$ and $\mA$ are the degree matrix in \eref{degree_mtx} and the adjacency matrix, respectively. In this paper, we find it more convenient to use the normalized Laplacian matrix 
\cite{chung_spectral_1997}, defined as
\begin{align*}
    \mL_{\text{norm}} & \bydef \mD^{-1/2} \mL_{\text{unnorm}} \mD^{-1/2} \\
    & = \mI - \mD^{-1/2}\mA \mD^{-1/2}.
\end{align*}
The choice of unnormalized or normalized Laplacian makes no essential difference to our analysis in \sref{Bound}. The latter is chosen because it leads to simpler expressions in some of our derivations. For notational simplicity, we will drop the subscript in $\mL_{\text{norm}}$, calling it $\mL$ in what follows.

Intuitively, the Laplacian matrix is analogous to the continuous Laplacian operator $-\nabla^2$ or $-\frac{d^2}{dy^2}$ on the real line. In fact, when the underlying graph is a line or a cycle, $\mL$ provides the standard stencil approximation for the second-order differentiation operator. The same holds for higher-dimensional lattices. In more general settings where the graphs are formed by sampling an underlying continuous manifold, the Laplacian matrix converges at high sampling densities to the Laplace-Beltrami operator, a differential geometric analogy to the second derivative \cite{belkin_problems_2003}. 


By construction, $\mL$ is a real symmetric matrix. We can therefore diagonalize $\mL$ as
\begin{equation}\label{eq:spectral_decomposition}
    \mL = \mF \mLambda \mF^T,
\end{equation}
where $\mF$ is an orthogonal matrix whose columns are the eigenvectors of $\mL$, and $\mLambda \bydef \diag\set{\lambda_1, \lambda_2, \ldots, \lambda_N}$
is a diagonal matrix of eigenvalues, which are all real. $\mL$ can be shown to be positive semidefinite with rank less than $N$, so we can order the eigenvalues as
$0 = \lambda_1 \le \lambda_2 \le \ldots \le \lambda_N$.

A large number of the topological properties of a graph can be inferred from the spectrum of its graph Laplacian \cite{chung_spectral_1997}.
For example, a graph is connected (meaning that a path can always be found connecting one vertex to the other) if and only if the smallest
eigenvalue ($\lambda_1 = 0$) has multiplicity one. The corresponding unit-norm eigenvector $\vf_1$ is defined by 
\begin{equation}\label{eq:dc}
f_1(v) = \sqrt{\frac{\deg(v)}{\sum_{u \in V} \deg(u)}},
\end{equation}
where $\deg(v)$ is the degree of the vertex $v$. One can also show that the maximum possible eigenvalue of $\mL$ is equal to $2$,
attained only by bipartite graphs. (These are graphs with two mutually exclusive subsets of vertices $U_0$ and $U_1$ such that every edge connects a vertex in $U_0$ to a vertex in $U_1$.)

Given a signal $\vx \in \ell^2(G)$, we can represent it in terms of the eigenvectors of $\mL$ by computing
\begin{equation}\label{eq:graph_Fourier}
 \widehat{\vx} = \mF^T \vx,
\end{equation}
where $\widehat{\vx}$ is called the \emph{graph Fourier transform} of $\vx$. The matrix $\mF^T$ represents
the Fourier transform operator%
\footnote{There may be eigenvalues of $\mL$ with multiplicity greater than one, so we should really think of the Fourier transform as the set of projections onto the eigenspaces associated with each unique eigenvalue. The Fourier transform defined in this way is unique up to unitary transformations within eigenspaces. We can choose an orthogonal basis in each eigenspace, ensuring that $F$ is orthogonal.}%
. Since $\mF$ is orthogonal, $\mF \mF^T = \mI$. It follows that we can invert the Fourier transform by taking
\begin{equation*}
 \vx = \mF \widehat{\vx}.
\end{equation*}

Using the Laplacian eigenvectors as a surrogate Fourier basis is a standard approach in the literature for defining signal processing operations on graphs \cite{coifman_diffusion_2006,coifman_diffusion_2006b, hammond_wavelets_2010,pesenson_sampling_2008, narang_perfect_2012}. It may not seem immediately obvious, though, that the analogy is a fair one. In what follows, we provide some justification for this approach. 


First, consider the special case of a cycle graph, illustrated in \fref{cyclegraph:1}. Signals defined on this graph can be thought of as discrete, periodic signals. The Laplacian of this graph is a circulant matrix, and is thus diagonalized by a discrete Fourier transform (DFT) matrix. Thus, in this case
the Laplacian eigenbasis is exactly the common sine/cosine DFT basis. \fref{cyclegraph:2} shows several such eigenvectors, which exhibit sinusoidal characteristics with increasing oscillation frequencies.



\begin{figure}[t]
\centering
\subfigure[]{\label{fig:cyclegraph:1}
\centering
\includegraphics[width=.2\textwidth]{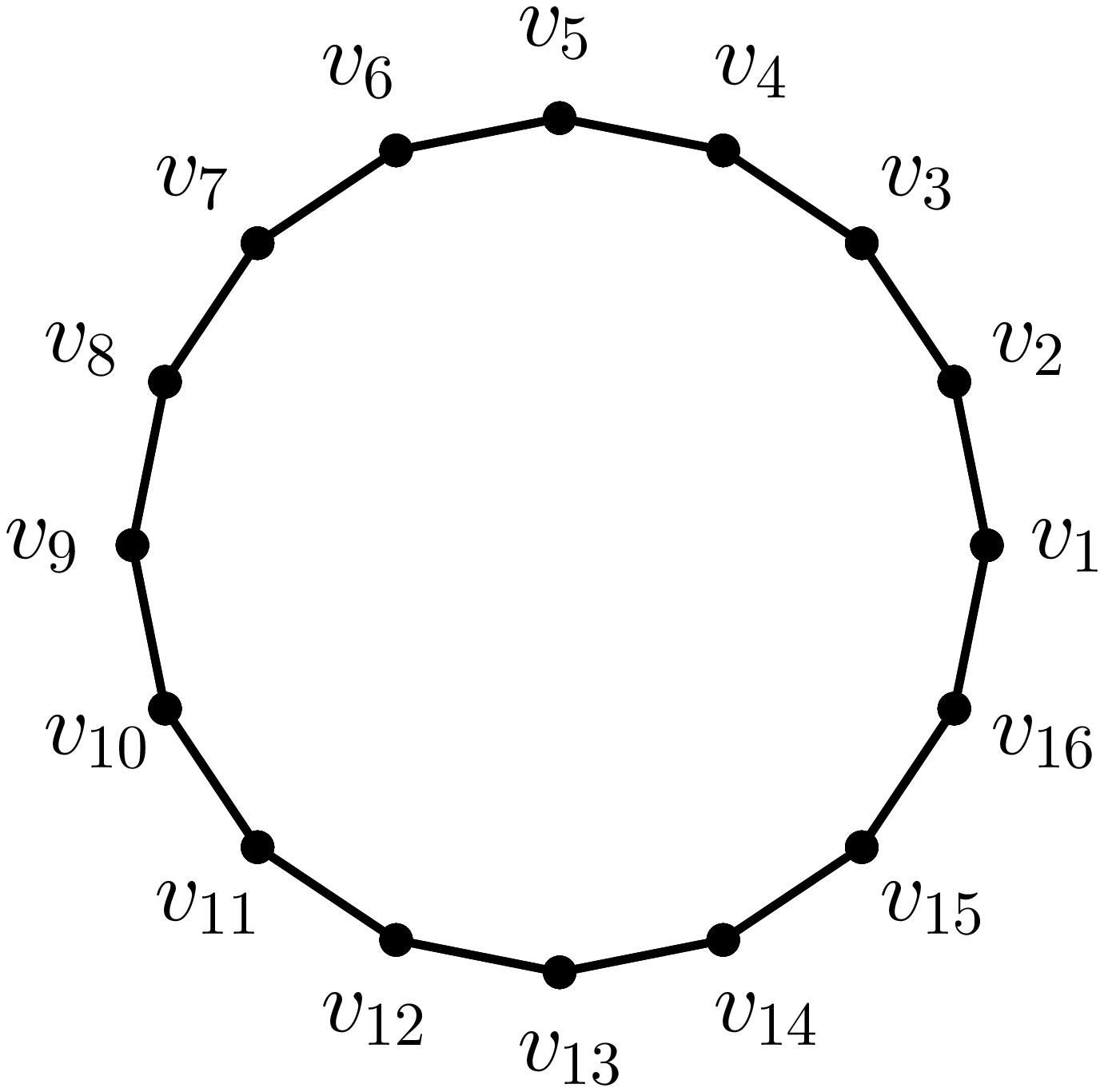}}
\hspace{8ex}
\subfigure[]{\label{fig:cyclegraph:2}
\centering
\includegraphics[width=.25\textwidth]{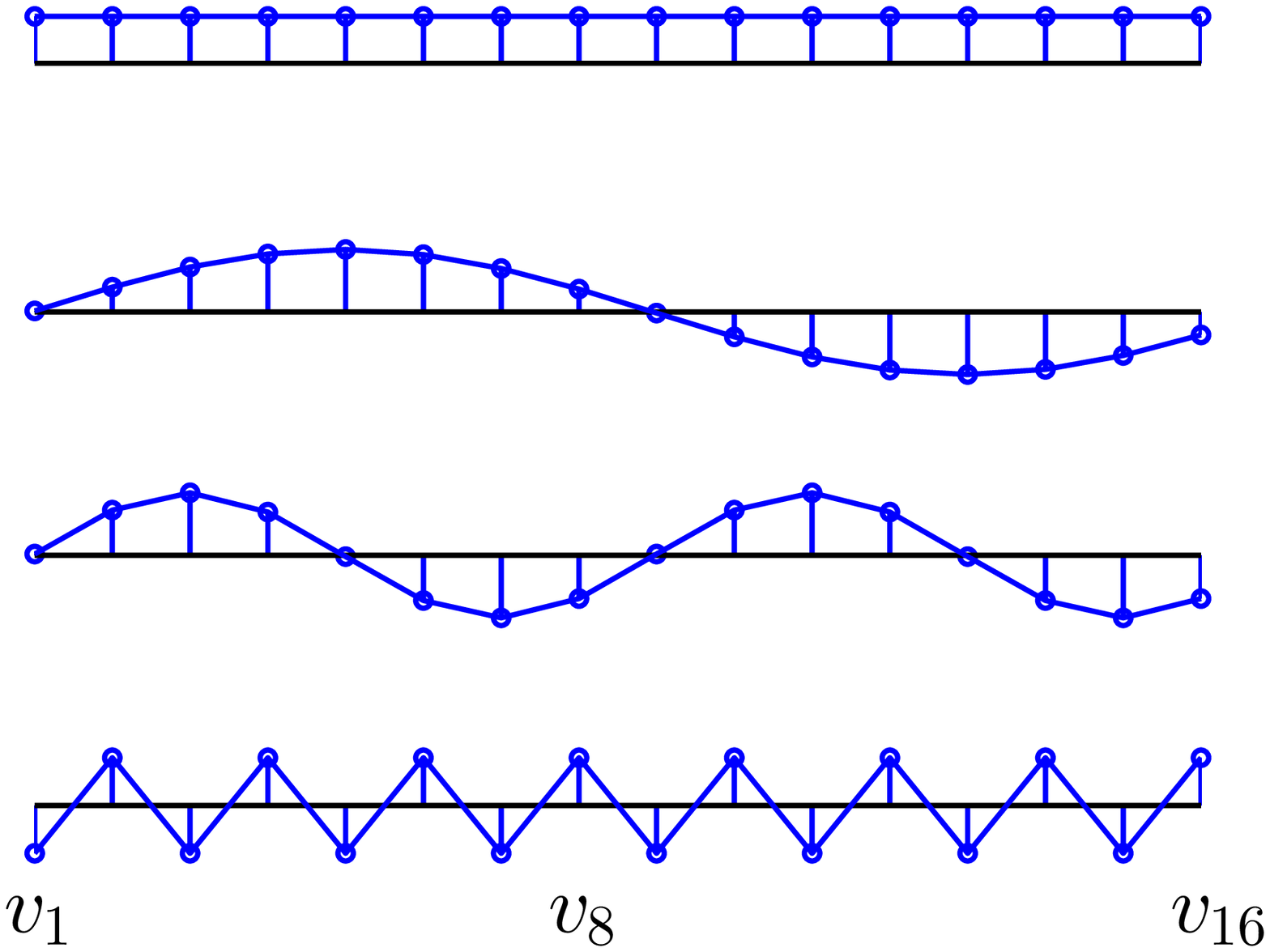} }
\caption{(a) A cycle graph with 16 vertices. Signals defined on this graph are equivalent to standard discrete, periodic signals. (b) Several eigenvectors of the graph Laplacian. These eigenvectors exhibit the sinusoidal characteristics
of the DFT basis.} 
\label{fig:cyclegraph}
\end{figure}

\begin{figure*}[t]
\centering
\input{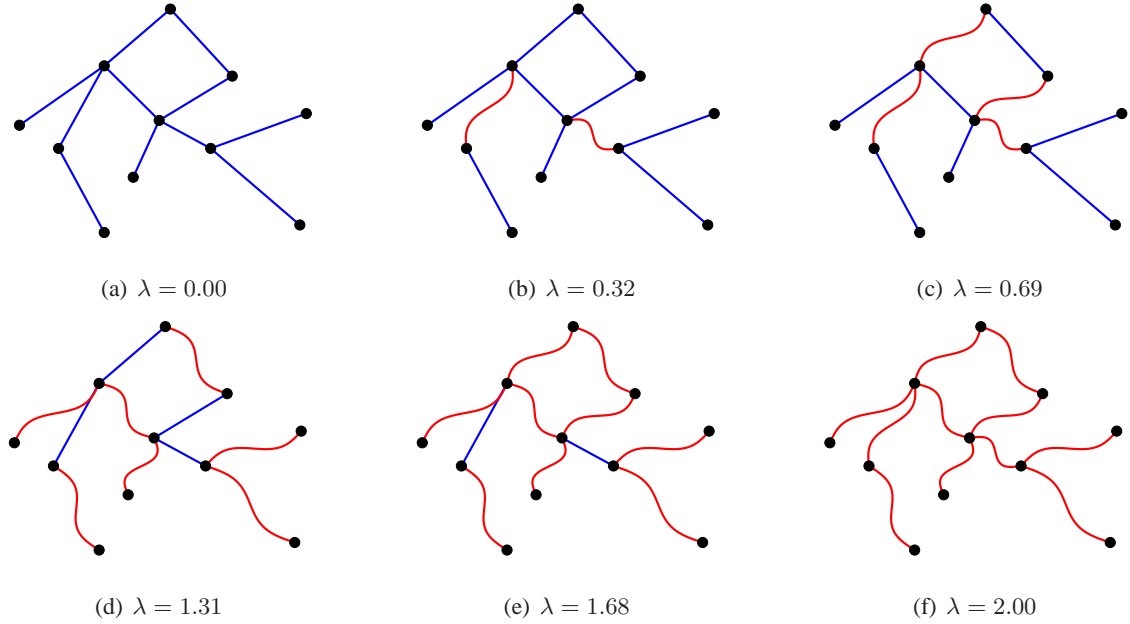}
\caption{Some Laplacian eigenvectors of a graph. Straight lines indicate that values on joined vertices have the same sign;
wavy lines indicate that there is a sign change between the joined vertices. As is evident, eigenvectors associated with larger eigenvalues correspond to more sign changes and thus
faster variation.}
\label{fig:eigvec}
\end{figure*}

For general graphs, of course, the Laplacian eigenbasis is no longer the DFT basis. Nonetheless, the eigenvectors still satisfy our intuition about frequency. For example, we would like to say that a signal is ``highpass'' if its value changes significantly between neighboring vertices, and that it is ``lowpass''  if its value varies very little. To quantify the variation of a signal on a graph, we can construct an $N \times M$ \emph{normalized incidence matrix} $\mS$ \cite{GodsilR:2001}, where each column of $\mS$ corresponds to an edge $e = (u,v)$ and has exactly two nonzero values: $+\frac{1}{\sqrt{\deg(u)}}$
in the row corresponding to vertex $u$, and $-\frac{1}{\sqrt{\deg(v)}}$ in the row corresponding to vertex $v$. The choice
of $(u,v)$ or $(v,u)$, and therefore the signs involved, is arbitrary for each edge (though it is important that each
column have one positive and one negative value.)  For any $\vx \in \ell^2(G)$, the vector $\vy = \mS^T \vx$ is a signal
on the edges of the graph, where each edge has the difference between the normalized values\footnote{The normalization by $\frac{1}{\sqrt{\deg(u)}}$ will limit the undue effect on the Laplacian of a vertex with a large number of incident edges.} of $\vx$ on its endpoint vertices. So, in a way, $\vy$ is the ``derivative'' of $\vx$. For any nonzero signal $\vx$, we can then measure its normalized variation on the graph as
\begin{align}
\frac{1}{\norm{\vx}^2} \sum_{u \sim v} \left(\frac{x(u)}{\sqrt{\deg(u)}} - \frac{x(v)}{\sqrt{\deg(v)}}\right)^2 & = \frac{1}{\norm{\vx}^2} \norm{\vy}^2 \nonumber \\
& = \frac{1}{\norm{\vx}^2}\vx^T \mS \mS^T \vx \nonumber \\
& = \frac{1}{\norm{\vx}^2}\vx^T \mL \vx,
\label{eq:norm_var}
\end{align}
where the last equality ($\mS \mS^T = \mL$) is well-known and easy to verify \cite{GodsilR:2001}. When the signal $\vx$ is the $i$th eigenvector $\vf_i$ of $\mL$, the normalized variation in \eref{norm_var} becomes $\lambda_i$, the corresponding eigenvalue. This justifies the usage of Laplacian eigenvalues as frequencies: eigenvectors corresponding to the higher eigenvalues of $\mL$ are the high-variation components, and the lower eigenvalues correspond to low-variation components. We illustrate this fact with an example in Figure \ref{fig:eigvec}.


\subsection{Graph and Spectral Spreads}
\label{sec:spreads}

We would like to quantify the localization of a signal on a graph in both
the graph and spectral domains. To do so, we look to the definitions of
analogous quantities in classical time-frequency analysis. For a nonzero signal $\vx \in \LL^2(\RR)$,
its time spread about a point $t_0$ is defined by \cite{VK_book}
\begin{equation}\label{eq:local_spread_classical}
    \Delta_{t,t_0}^2 \bydef \frac{1}{\norm{\vx}^2} \int_{-\infty}^\infty (t - t_0)^2 |x(t)|^2 dt.
\end{equation}
The overall time spread of $x(t)$ is then obtained by minimizing over $t_0$, \emph{i.e.},
\begin{equation}\label{eq:global_spread_classical}
    \Delta_t^2 \bydef \min_{t_0} \frac{1}{\norm{\vx}^2} \int_{-\infty}^\infty (t - t_0)^2 |x(t)|^2 dt,
\end{equation}
where the minimizing value of $t_0$ is given by $t_0 = \frac{1}{\norm{\vx}^2} \int_{-\infty}^\infty t |x(t)|^2 dt$. Generalizing \eref{local_spread_classical} to signals defined on graphs, we introduce the following definition \cite{agaskar_uncertainty_2011,agaskar_uncertainty_2012}.

\begin{definition}[Graph spread] For a nonzero signal $\vx \in \ell^2(G)$, its graph spread about a vertex $u_0$ is\begin{align}
    \label{eq:local_spread}
    \Delta_{g,u_0}^2(\vx) & \bydef \frac{1}{\norm{\vx}^2} \sum_{v \in V} d(u_0,v)^2 x(v)^2 \\
                     & = \frac{1}{\norm{\vx}^2} \, \vx^T \mP_{u_0}^2 \vx, \nonumber
\end{align}
where $d(\cdot, \cdot)$ is the distance metric described in Section \ref{subsec:graphdefs}, and $\mP_{u_0}$ is a diagonal matrix defined as
\begin{equation}
\label{eq:Pu}
\mP_{u_0} \bydef \diag\left\{d(u_0, v_1), d(u_0, v_2), \ldots, d(u_0, v_N)\right\}.
\end{equation}
\end{definition}

\begin{remark}
Similar to \eref{global_spread_classical}, we can also define the overall (\emph{i.e.}, global) graph spread of $\vx$ as
\begin{equation}
    \label{eq:global_spread}
    \Delta_g^2(\vx) \bydef \min_{u_0 \in V} \frac{1}{\norm{\vx}^2} \, \vx^T \mP_{u_0}^2 \vx.
\end{equation}
For our subsequent analysis on uncertainty principles though,\label{page:global_spread} we will focus on the local graph spread (\emph{i.e.}, about
a particular vertex $u_0$) as defined in \eref{local_spread}. Unlike classical domains such as the real line whose topology is shift-invariant, the ``landscape'' of a graph can look very different around different vertices. Thus, it is important to explicitly specify the center vertex $u_0$ 
when considering the graph spread and uncertainty principles. If needed, global versions can always be obtained through finite minimization over all $u_0 \in V$.
\end{remark}


The spectral spread of a signal defined on graphs requires more thought. In the classical case, the frequency spread of a real-valued signal $x(t) \in \LL^2(\RR)$ is given by \cite{VK_book}
\begin{equation}\label{eq:spectral_spread_classical}
 \Delta_\omega^2 \bydef \frac{1}{\norm{\vx}^2} \int_{-\infty}^\infty \omega^2 |\widehat{x}(\omega)|^2 \frac{d \omega}{2 \pi},
\end{equation}
where $\widehat{x}(\omega)$ is the Fourier transform of $x(t)$.
This expression is simpler than that of the time spread in \eref{local_spread_classical} because the frequency center is chosen to be $\omega_0 = 0$ due to the symmetry of the Fourier transforms of real-valued signals. On recognizing that $\omega^2 \widehat{x}(\omega)$ is the Fourier transform of $\frac{-d^2}{dt^2} x(t)$ and using Parseval's identity, we can rewrite \eref{spectral_spread_classical} as
\begin{equation}\label{eq:spectral_spread_classical2}
\Delta_\omega^2 = \frac{1}{\norm{\vx}^2} \int_{-\infty}^\infty x(t) \frac{-d^2}{dt^2} x(t) dt.
\end{equation}
Generalizing to the graph case, treating $\mL$ as analogous to the operator $-\frac{d^2}{dt^2}$, we obtain the following definition \cite{agaskar_uncertainty_2011,agaskar_uncertainty_2012}.
\begin{definition}[Spectral spread]
For a nonzero signal $\vx \in \ell^2(G)$, we define its spectral spread as
\begin{align}
 \Delta_s^2(\vx) &\bydef \frac{1}{\norm{\vx}^2} \, \vx^T \mL \vx  \label{eq:spectral_spread}\\
 	&= \frac{1}{\norm{\vx}^2} \sum_{n=1}^N \lambda_n \, \abs{\widehat{x}_n}^2, \label{eq:spectral_spread2}
\end{align}
where the second equality follows from the decomposition of $\mL$ in \eref{spectral_decomposition} and the definition of graph Fourier transforms in \eref{graph_Fourier}.
\end{definition}

\begin{remark}
The equivalent definitions in \eref{spectral_spread} and \eref{spectral_spread2} reveal two different facets of the spectral spread: while \eref{spectral_spread2} perhaps more clearly justifies the ``spectral'' nature of $\Delta_s^2(\vx)$, the form in \eref{spectral_spread} shows that $\Delta_s^2(\vx)$ can also be understood as the normalized variation of $\vx$ introduced in \eref{norm_var}.
\end{remark}


\section{Uncertainty Principles: Bounds and Characterizations}
\label{sec:Bound}

Intuitively, we can reason that there should exist a tradeoff between the graph and spectral spreads of a signal. If the graph spread $\Delta_g^2$ is small, then the signal must resemble an impulse centered at some vertex; in this case, the normalized variation (\emph{i.e.}, the spectral spread $\Delta_s^2$) should be high. If instead $\Delta_s^2$ is small, then the signal cannot vary too quickly; it will thus take a long distance for the signal values to drop significantly from the peak value, in which case the graph spread will be high. How can one precisely quantify the above intuition? What are the signals with a given spectral spread that are maximally localized on the graph?
These are the fundamental questions addressed in this section.

\subsection{The Feasibility Region}

In the classical uncertainty principle, not all pairs of time-frequency spreads $(\Delta_t^2, \Delta_{\omega}^2)$ are achievable, and the tradeoff is quantified by the celebrated inequality $\Delta_t^2 \Delta_{\omega}^2 \geq \frac{1}{4}$, which holds for any nonzero function $x(t) \in \LL^2(\RR)$ \cite{folland_uncertainty_1997, VK_book}. Furthermore, this bound is tight. In fact, any pair of the form $(\Delta_t^2, \Delta_\omega^2) = (c, \frac{1}{4c})$ for $c > 0$ is achievable by a function of the form
$x(t) = \exp\left(-\frac{t^2}{4 c}\right)$.


In a similar way, we are interested in characterizing the following \emph{feasibility region} 
\begin{align}\label{eq:feasible}
\DD \bydef  \{ (s, g): & \, \DS(\vx) = s, \, \DG(\vx) = g \nonumber \\ & \text{ for some nonzero }\vx \in \ell^2(G)\},
\end{align}
containing all pairs of the form $(\DS, \DG)$ that are achievable on a graph $G$, using $u_0$ as the  center vertex.

\begin{proposition}\label{prop:feasible_props}
Let $\DD$ be the feasibility region for a \emph{connected} graph $G$ with $N$ vertices. Then the following properties hold:
\begin{enumerate}
    \item[(a)] $\DD$ is a closed subset of $[0, \lambda_{N}] \times [0, \mathcal{E}_G^2(u_0)]$, where $\lambda_{N} \leq 2$ is the largest eigenvalue of graph Laplacian $\mL$, and $\mathcal{E}_G(u_0) \bydef \max_{v \in V} d(u_0, v)$ is the \emph{eccentricity} of the center vertex $u_0$.

\item[(b)] $\DD$ intersects the horizontal axis at exactly one point, $(1, 0)$, and the vertical axis at exactly one point, $(0, \vf_1^T \mP_{u_0}^2 \vf_1)$, where $\vf_1$ is the eigenvector defined in \eref{dc}.

\item[(c)] The points $(1, \mathcal{E}_G^2(u_0))$ and $(\lambda_{N}, \vf_N^T \mP_{u_0}^2 \vf_N)$, where $\vf_N$ is any unit-norm eigenvector associated with $\lambda_{N}$, belong to $\DD$.

\item[(d)] $\DD$ is a convex set if the number of vertices $N \geq 3$.
\end{enumerate}
\end{proposition}
\begin{IEEEproof}
    (a) The graph and spectral spreads of any nonzero signal can be bounded by the largest and smallest eigenvalues of $\mL$ and $\mP_{u_0}^2$. More precisely, using the Rayleigh inequalities \cite{lancaster_theory_1985}, we have
        \begin{equation*}
            0 = \lambda_{1} \leq \frac{\vx^T \mL \vx}{\vx^T \vx} \leq \lambda_{N}
        \end{equation*}
        and, similarly, 
        \begin{equation*}
            0 \leq \frac{\vx^T \mP_{u_0}^2 \vx}{\vx^T \vx} \leq \max_{1 \le i \le N} 
            (\mP_{u_0}^2)_{ii} = \mathcal{E}_G^2(u_0).
        \end{equation*}
        $\DD$ is compact, and therefore closed, because it is the image of a compact set under a continuous transform \cite{strichartz_way_2000}.
Specifically, if we take the
unit sphere in $\RR^N$, a compact set, and apply the map $f : \vx \mapsto (\DS(\vx), \DG(\vx))$, which is continuous on the unit sphere, we get the whole uncertainty region.

(b) A signal has zero graph spread (\emph{i.e.}, $\Delta_{g, u_0}^2(\vx) = 0$) if and only if it is an impulse supported on $u_0$, \emph{i.e.}, $x(v) = c$ if $v = u_0$ and $x(v) = 0$ otherwise, for some nonzero scalar $c$. Meanwhile, using \eref{spectral_spread} and \eref{norm_var}, one can verify that the normalized variation (and thus the spectral spread $\Delta_s^2$) of such impulse signals is equal to $1$. It follows that $(1, 0)$ is the only point that lies at the intersection of $\DD$ and the horizontal axis. Next, consider the intersection of $\DD$ with the vertical axis. Since $\Delta_s^2(\vx) = \vx^T \mL \vx / \norm{\vx}^2 \ge \lambda_1 = 0$, the spectral spread $\Delta_s^2(\vx) = 0$ if and only if $\vx$ is an eigenvector of $\mL$ associated with the smallest eigenvalue $\lambda_1 = 0$. (See \eref{dc} for an example.) Such eigenvectors are also unique (up to scalar multiplications) since the smallest eigenvalue $\lambda_1$ of connected graphs always has multiplicity one \cite{chung_spectral_1997}. 

(c) The inclusion of $(\lambda_{N}, \vf_N^T \mP_{u_0}^2 \vf_N)$ in $\DD$ is clear. For the first point $(1, \mathcal{E}_G^2(u_0))$, consider an impulse function supported at the furthest vertex on the graph from $u_0$. Similar to (b), we can compute its spectral and graph spreads as $\Delta_s^2 = 1$ and $\Delta_{g, u_0}^2 = \mathcal{E}_G^2(u_0)$, respectively.

(d) See Appendix~\ref{appendix:convexity}. 
\end{IEEEproof}

\begin{remark}
\fref{feasible} illustrates a typical feasibility region $\DD$ as specified by Proposition \ref{prop:feasible_props}. The boundedness and convexity of $\DD$ imply that the entire region can be completely characterized by its upper and lower boundaries: any pair between the two boundaries must also be achievable. Furthermore, the lower boundary must be convex and the upper boundary must be concave.
\end{remark}

\begin{figure}[t]
\centering
\input{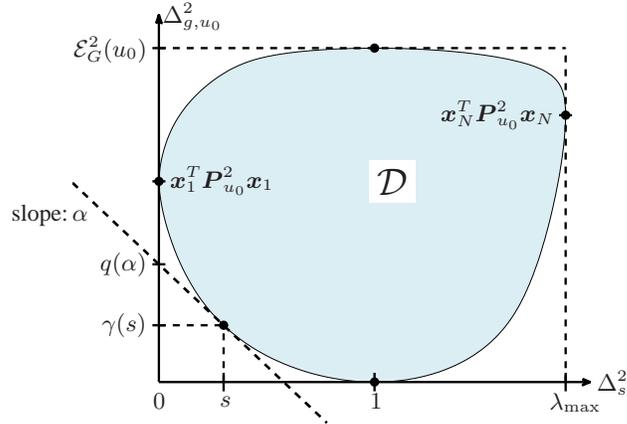}
\caption{The feasibility region $\DD$ for the spectral and graph spreads. $\DD$ is a bounded and convex set that intersects the horizontal (and vertical) axis at exactly one point. The lower boundary of $\DD$ can be implicitly computed by considering supporting lines of varying slopes. The achievable region must lie in the half-plane above
the supporting line (found by solving an eigenvalue problem.)}
\label{fig:feasible}
\end{figure}

\subsection{The Uncertainty Curve}
\label{sec:gamma_s}

In what follows, we will describe a technique for computing the lower boundary curve of $\DD$, which we call the \emph{uncertainty curve}. 
\begin{definition}
Given a connected graph $G$, the uncertainty curve with respect to a center vertex $u_0$ is
\begin{align}
\gu(s) & \bydef  \min_{\vx} \, \DG(\vx) \,\text{ subject to } \, \DS(\vx) = s \nonumber\\
& = \min_{\vx} \, \vx^T \mP_{u_0}^2 \vx \, \text{ subject to }\, \vx^T \vx = 1 \text{ and } \vx^T \mL \vx = s,\label{eq:qcqp}
\end{align}
for all $s \in [0, \lambda_{N}]$.
\end{definition}

\begin{remark}
    We could also define and study the upper boundary curve of $\DD$ in a similar way. We choose to focus on the lower boundary curve because it 
    provides an uncertainty bound analogous to the classical bound (\ref{eq:classical_uncertainty}). 
    We will say that a signal $\vx$ achieves \label{page:achieves_curve} the uncertainty curve if $\Delta_{g,u_0}^2(\vx) = \gu(\Delta_s(\vx)^2)$.
\end{remark}

We note that \eref{qcqp} is a quadratically constrained quadratic program \cite{boyd_convex_2004}. 
The equality constraints make the problem nonconvex. On differentiating the corresponding Lagrangian function
\begin{equation*}
    \Lambda(\vx; \alpha, \lambda) \bydef \vx^T \mP_{u_0}^2 \vx - \alpha (\vx^T \mL \vx - s) - \lambda (\vx^T \vx - 1),
\end{equation*}
we see that the optimal solution $\vx^\ast$ to \eref{qcqp} must satisfy
\begin{equation*}
    \label{eq:nec_cond}
  (\mP_{u_0}^2 - \alpha \mL) \vx^\ast = \lambda \vx^\ast
\end{equation*}
for some $\alpha, \lambda \in \RR$. If we treat $\alpha$ as being fixed, then the above equality becomes an eigenvalue problem. This observation leads us to study the matrix-valued function
\begin{equation}\label{eq:M_alpha}
\mM(\alpha) \bydef \mP_{u_0}^2 - \alpha \mL.
\end{equation}
For any $\alpha$, the smallest eigenvalue of $\mM(\alpha)$, denoted by 
\begin{equation*}
q(\alpha) \bydef \lambda_{\min}(\mM(\alpha)),
\end{equation*}
and its associated eigenspace, denoted by $\Ss(\alpha)$, are key to our analysis of the uncertainty curve $\gu(s)$.
\begin{proposition}\label{prop:gen_point}
For any $\alpha \in \RR$ and any unit-norm eigenvector $\vv$ in $\Ss(\alpha)$, the point $(\vv^T \mL \vv, \vv^T \mP_{u_0}^2 \vv)$ is on $\gu(s)$.
\end{proposition}
\begin{IEEEproof}
Let $\vx$ be an arbitrary signal with $\norm{\vx} = 1$. By definition, $\DG(\vx) - \alpha \DS(\vx) = \vx^T \mM(\alpha) \vx$. Applying Rayleigh's inequality to $\mM(\alpha)$ thus leads to
\begin{align}
    \DG(\vx) - \alpha \DS(\vx) &\ge q(\alpha)\label{eq:halfplane}\\
			&= \vv^T \mP_{u_0}^2 \vv - \alpha \vv^T \mL \vv, \label{eq:eigen_v}
\end{align}
where \eref{eigen_v} comes from the fact that $\vv$ is an eigenvector associated with $q(\alpha)$. Let $s = \vv^T \mL \vv$. On specializing the relationship \eref{eigen_v} to those signals $\vx$ satisfying $\DS(\vx) = s$, we have
\[
\DG(\vx) \ge \vv^T \mP_{u_0}^2 \vv,
\]
which indicates that the point $(\vv^T \mL \vv, \vv^T \mP_{u_0}^2 \vv)$ must lie on the uncertainty curve $\gu(s)$.
\end{IEEEproof}

There is an interesting geometric interpretation of the above derivations: as illustrated in \fref{feasible}, 
for any $\alpha$, the inequality in \eref{halfplane} defines a half-plane in which $\DD$ must lie. 
The boundary of the half-plane, a line of slope $\alpha$ defined by
\[
\DG - \alpha \DS = q(\alpha),
\]
provides a tight lower bound to $\DD$. Varying the values of $\alpha$ generates a family of such half-planes, the intersection of which contains $\DD$.
For readers familiar with convex analysis, we note that $q(\alpha)$ is the Legendre transform of $\gu(s)$ \cite{boyd_convex_2004}.

Proposition~\ref{prop:gen_point} guarantees that any nonzero eigenvector of $\mM(\alpha)$ associated with the smallest eigenvalue $q(\alpha)$ generates a point on the curve $\gu(s)$. Next, we will show that the converse is also true: every point on $\gu(s)$ is achieved by an eigenvector in $\Ss(\alpha)$ for some $\alpha$. To establish this result, we need to introduce the following two functions:
\begin{equation}\label{eq:hpm}
h_{+}(\alpha) \bydef \max_{\vx \in \Ss(\alpha): \,\norm{\vx} = 1} \vx^T \mL \vx \quad \text{and}\quad h_{-}(\alpha) \bydef \min_{\vx \in \Ss(\alpha): \,\norm{\vx} = 1} \vx^T \mL \vx,
\end{equation}
which measure, respectively, the maximum and minimum spectral spread (\emph{i.e.}, the horizontal coordinate on the $s$--$g$ plane) that can be achieved by eigenvectors in $\Ss(\alpha)$.

%


\begin{lemma}
    \label{lem:hprops}
    The following properties hold for $h_{+}(\alpha)$ and $h_{-}(\alpha)$.
    \begin{enumerate}
        \item[(a)] They are increasing functions, \emph{i.e.}, $h_{+}(\alpha_1) \le h_{+}(\alpha_2)$ and $h_{-}(\alpha_1) \le h_{-}(\alpha_2)$ for all $\alpha_1 < \alpha_2$.
        
        \item[(b)] They have the same limits as $\abs{\alpha}$ tends to infinity:
    \begin{align}	
	\lim_{\alpha \rightarrow -\infty} h_+(\alpha) & = \lim_{\alpha \rightarrow -\infty} h_-(\alpha) = 0, \label{eq:limit_left}\\ 
	\shortintertext{and}
	\lim_{\alpha \rightarrow +\infty} h_+(\alpha) & = \lim_{\alpha \rightarrow +\infty} h_-(\alpha) = \lambda_{N}.\label{eq:limit_right}
	\end{align}
        
\item[(c)] \label{page:hprops_c} On any finite interval $[a, b]$, the functions $h_{+}(\alpha)$ and $h_{-}(\alpha)$ differ 
            on at most a finite number of points, denoted by $\mathcal{B} \bydef \set{\beta_1, \beta_2, \ldots, \beta_k}$ 
            for some $k \ge 0$. Except for these points, $h_{+}(\alpha)$ and $h_{-}(\alpha)$ coincide, are continuous, and satisfy
        \begin{equation}\label{eq:hpm_derivative}
        h_{+}(\alpha) = h_{-}(\alpha) = -q'(\alpha), \qquad \text{for all } \alpha \in [a, b] \setminus \mathcal{B},
        \end{equation}
        where $q'(\alpha)$ is the derivative of $q(\alpha)$. At the points, if any, where they do differ, $h_+(\alpha)$ and $h_{-}(\alpha)$
        have jump discontinuities. Moreover, for all $\beta \in \mathcal{B}$,
        \[
        h_+(\beta) = \lim_{\alpha \rightarrow \beta^+} h_+(\alpha)  > \lim_{\alpha \rightarrow \beta^-} h_-(\alpha) 
        = h_{-}(\beta),
        \]
        where the limits are taken as $\alpha$ approaches $\beta$ from the positive and negative sides, respectively.
    \end{enumerate}
\end{lemma}
\newcounter{qprops}
\setcounter{qprops}{\value{lemma}}
\label{sec:lem:qprops}
\begin{IEEEproof}
    See Appendix~\ref{appendix:diff_analysis}.
\end{IEEEproof}

The results of Lemma~\ref{lem:hprops} are illustrated in \fref{alpha_s_and_gamma}(a), where we plot a typical example of $h_{+}(\alpha)$ and $h_{-}(\alpha)$:
as $\alpha$ increases from $-\infty$ to $+\infty$, the values of the functions increase from $0$ to $\lambda_{N}$. 
Within any finite interval, $h_{+}(\alpha) = h_{-}(\alpha)$ except at a finite number of points (\emph{e.g.}, the point $\beta$ in the figure). At these ``jump points'', $h_{+}(\alpha)$ is right-continuous, whereas $h_{-}(\alpha)$ is left-continuous. 

\begin{figure*}[t]
    \centering
    \psfrag{000}[r][r]{\scriptsize$0$}
    \psfrag{AL}[l][l]{\scriptsize$\alpha$}
    \psfrag{SAL}[][]{\scriptsize$h_+(\alpha)$ and $h_-(\alpha)$}
    \psfrag{LMAX}[l][l]{\scriptsize$\lambda_{N}$}
    \psfrag{SSS}[Bl][Bl]{\scriptsize$s$}
    \psfrag{GS}[r][r]{\scriptsize$\gu(s)$}
    \psfrag{QA}[r][r]{\scriptsize$q(\alpha)$}
    \psfrag{SLP}{\scriptsize slope: $\alpha$}
    \psfrag{HMB}{\scriptsize $h_-(\beta)$}
    \psfrag{HPB}{\scriptsize $h_+(\beta)$}
    \psfrag{HA}{\scriptsize $h_+(a) = h_-(a)$}
    \psfrag{HB}{\scriptsize $h_+(b) = h_-(b)$}
    \psfrag{BETA}{\scriptsize $\beta$}
    \psfrag{PTA}{\scriptsize $a$}
    \psfrag{PTB}{\scriptsize $b$}
    \psfrag{AAA}{(a)}
    \psfrag{BBB}{(b)}
    \includegraphics[width=0.8\textwidth,height=.45\textwidth]{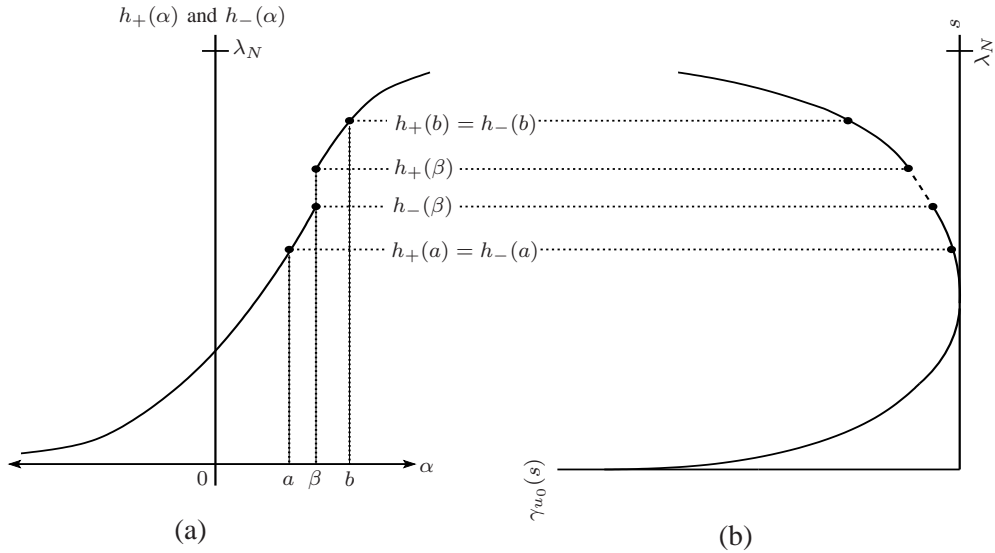}
    \caption{The mapping of the eigenvectors in
    $S(\alpha)$ onto the $s$--$g$ plane is shown. In (a), $h_+(\alpha)$ and $h_-(\alpha)$
    are plotted against $\alpha$ (they coincide except at jumps in the plot.) 
    They are, respectively, the maximum and minimum spectral spreads of elements of the eigenspace $S(\alpha)$.
    Any element of $S(\alpha)$ determines a point on the graph of $\gu(s)$. When $S(\alpha)$ is of dimension
    greater than one, it corresponds to a line segment on $\gu(s)$.}
    \label{fig:alpha_s_and_gamma}
\end{figure*}

Since we are only considering connected graphs, $\lambda_1 = 0$ has multiplicity 1, and so $\vf_1$ is the unique vector (up to scaling) that
achieves the uncertainty curve with $\DS = 0$. At the other end, $\lambda_N$ may have multiplicity, but some vector in its eigenspace
will achieve the uncertainty curve with $\DS = \lambda_N$. For values of $s \in (0, \lambda_{\max})$, we can use the following theorem
to precisely characterize vectors that achieve the uncertainty curve at $s$.
\begin{theorem}\label{thm:smallesteig}
    A signal $\vx \in \ell^2(G)$ with $\DS(\vx) \in (0, \lambda_{\max})$ achieves the uncertainty curve, \emph{i.e.}, $\DG(\vx) = \gamma(\DS(\vx))$,
    \emph{if and only if} it is a nonzero eigenvector in $\Ss(\alpha)$ for some $\alpha$.
\end{theorem}
\begin{IEEEproof}
The ``if'' direction has been established in Proposition \ref{prop:gen_point}. To prove the ``only if'' direction, we will show that for any 
signal $\vx \in \ell^2(G)$ that achieves the uncertainty curve, there is an $\alpha$ and a unit-norm eigenvector $\vv \in \Ss(\alpha)$ such that $\vv^T \mL \vv = \DS(\vx)$. 
Since both $\vx$ and $\vv$ lie on the uncertainty curve (with the former given as an assumption and the latter guaranteed by Proposition~\ref{prop:gen_point}), we have  $\DG(\vx) = \vv^T \mP^2_{u_0} \vv$, and thus
\[
\frac{1}{\norm{\vx}^2} \, \vx^T \mM(\alpha) \vx = \DG(\vx) - \alpha \DS(\vx) =  \vv^T \mM(\alpha) \vv = q(\alpha).
\]
Now, since $q(\alpha)$ is the smallest eigenvalue of $\mM(\alpha)$, the equality above implies that $\vx$ must also be an eigenvector associated with $q(\alpha)$. In fact, $\vx$ will be equal to $\vv$ (up to a scalar multiple) if $q(\alpha)$ has multiplicity one. The remainder of the proof verifies the claim, namely, for any $s \in (0,  \lambda_{N})$ we can find an $\alpha$ and a unit-norm eigenvector $\vv \in \Ss(\alpha)$ such that $\vv^T \mL \vv = s$.   
    
By part (b) of Lemma~\ref{lem:hprops}, we can always find some $a < b$ such that $h_{-}(a) < s < h_{+}(b)$. 
Furthermore, part (c) of Lemma~\ref{lem:hprops} ensures that, within the interval $[a, b]$, the two functions $h_{+}(\alpha)$ and $h_{-}(\alpha)$ differ 
(and are discontinuous) on at most a finite number of points. For notational simplicity, and without loss of generality, we assume that there 
is only one such discontinuity point, denoted by $\beta \in [a, b]$. As shown in \fref{alpha_s_and_gamma}, the interval $[h_{-}(a), h_{+}(b)]$ can now be written as the union of three subintervals
\[
[h_{-}(a), h_{-}(\beta)), \ [h_{-}(\beta), h_{+}(\beta)], \text{ and } (h_{+}(\beta), h_{+}(b)],
\]
to one of which $s$ must belong.

We first consider the case where $s \in [h_{-}(a), h_{-}(\beta))$. Lemma~\ref{lem:hprops} says that $h_{-}(\alpha)$ is a continuous function on 
$[a,\beta]$. By the intermediate value theorem, 
there exists some $\alpha_0  \in [a, \beta]$ such that $h_{-}(\alpha_0) = s$. By definition, 
$h_{-}(\alpha_0) = \min_{\vz \in \Ss(\alpha_0): \,\norm{\vz} = 1} \vz^T \mL \vz$. 
Since the eigenspace $\Ss(\alpha_0)$ has finite dimensions, the minimization can always be achieved by some unit-norm 
eigenvector $\vv \in \Ss(\alpha_0)$, \emph{i.e.}, $s = h_{-}(\alpha_0) = \vv^T \mL \vv$. The same line of 
reasoning can be used when $s$ belongs to the third subinterval, $(h_{+}(\beta), h_{+}(b)]$.
    This leaves us with the remaining case when $s \in [h_{-}(\beta), h_{+}(\beta)]$. Let
\[
\vv_+ \bydef \underset{\vz \in \Ss(\beta): \norm{\vz} = 1}{\operatorname{arg max}} \vz^T \mL \vz \quad \text{ and } \quad
\vv_- \bydef \underset{\vz \in \Ss(\beta): \norm{\vz} = 1}{\operatorname{arg min}} \vz^T \mL \vz,
\]
and consider the vector-valued function $\vy(\theta) \bydef \frac{\cos(\theta) \vv_+ + \sin(\theta) \vv_-}{1 + \sin(2 \theta)\vv_+^T\vv_-}$,
defined for $\theta \in [0, \pi/2]$.
The denominator is nonzero for every $\theta$, since $\vv_- \neq -\vv_+$ [otherwise we would have $h_-(\beta) = h_+(\beta)$].
So $\vy(\theta)$ is of unit norm and is a continuous function of $\theta$.
It also must belong to $\Ss(\beta)$ since it is a linear combination of two elements of the subspace. Furthermore, $\vy(0)^T \mL \vy(0) = h_+(\beta)$ and $\vy(\pi/2)^T \mL \vy(\pi/2) = h_-(\beta)$. By the intermediate value theorem, $\vy(\theta)$ for $\theta \in [0, \pi/2]$ achieves all the values in between. In particular, there exists some $\theta_0$ such that $\vy(\theta_0)^T \mL \vy(\theta_0) = s$. We note that since every element of $\Ss(\beta)$ achieves a point
on the line $g - \beta s = q(\beta)$, this interpolation procedure amounts to including the straight line segment between
the two endpoints as part of the uncertainty curve.
\end{IEEEproof}

\textit{Remark:} If $\Ss(\alpha)$ is one-dimensional for every $\alpha \in [a,b]$, or more generally if there is a single distinct eigenvalue function that achieves the minimum 
on $[a,b]$, then from Theorem \ref{thm:smallesteig} as well as
Lemma \ref{lem:hprops} and its proof, $q(\alpha)$ is analytic on $[a,b]$ and the corresponding portion of the uncertainty curve can be expressed in parametric form  as
\begin{equation}
    \begin{cases}
        s(\alpha) = -q'(\alpha) \\ 
        \gu(s) = q(\alpha) - \alpha q'(\alpha),
    \end{cases}
    \label{eq:gammas_parametric}
\end{equation}
where the first equality is due to (\ref{eq:hpm_derivative}) and the second is due to the fact that any vector in $\calS(\alpha)$ must
achieve a point on the line $g - \alpha s = q(\alpha)$.

In general, Theorem \ref{thm:smallesteig} and its proof justify a way to obtain the uncertainty curve: for every $\alpha$, we find 
the eigenvectors associated with the smallest eigenvalue of $\mM(\alpha)$. These eigenvectors will give us points on $\gu(s)$. By ``sweeping'' the values of $\alpha$ from $-\infty$ to $\infty$, the entire curve can then be traced.

\subsection{Fast Approximation Algorithm}
\label{sec:curve_approx}
In practice, of course, we must sample and work with a finite set of $\alpha$'s, which lead to an approximation of the true curve. In what follows, we describe an efficient algorithm that can compute an approximation---more specifically, an upper and lower bound---of $\gu(s)$ with any desired accuracy.

%
%

\begin{figure}
	\centering
    	\psfrag{UU}{\scriptsize Upper Bound}
    	\psfrag{TT}{\scriptsize True $\gu(s)$}
    	\psfrag{LL}{\scriptsize Lower Bound}
    	\psfrag{SS}{\scriptsize $\DS$}
    	\psfrag{GG}{\scriptsize $\DG$}
        \psfrag{A}[l][l]{\scriptsize $A$}
        \psfrag{B}[bl][bl]{\scriptsize $B$}
        \psfrag{C}[l][l]{\scriptsize $C$}
        \psfrag{D}[bl][bl]{\scriptsize $D$}
        \psfrag{E}[l][l]{\scriptsize $E$}
        \psfrag{F}[bl][bl]{\scriptsize $F$}
    	\subfigure[]{\label{fig:sandwich:1}
    		\centering
    		\includegraphics[width=.36\textwidth]{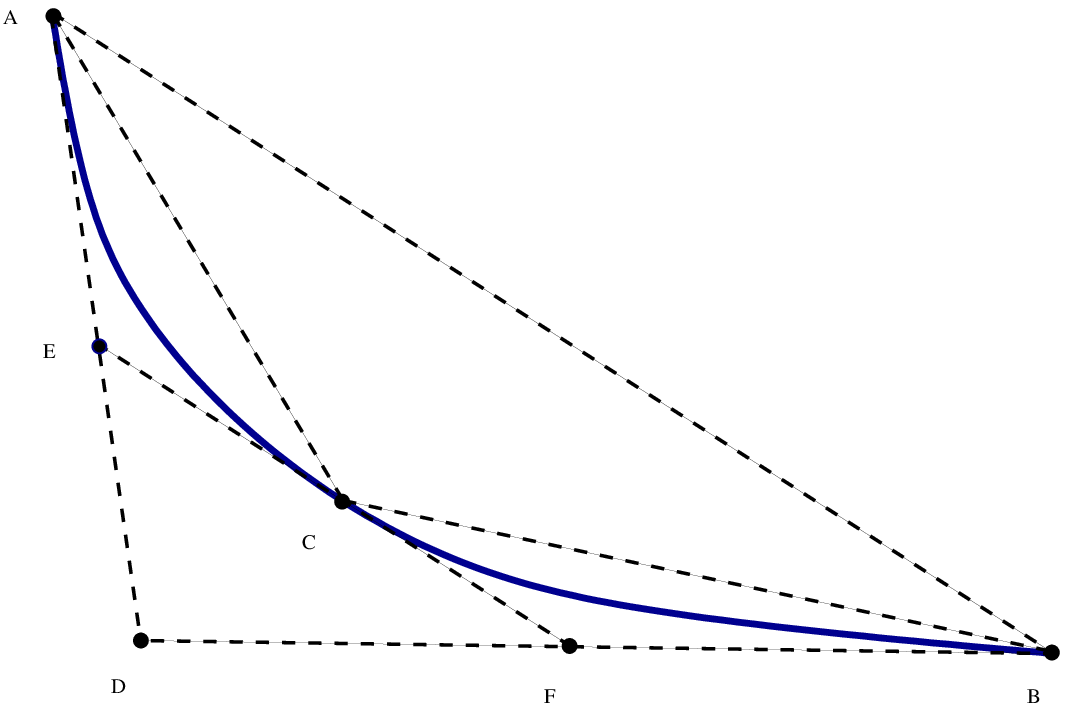}
   	    }
	\hspace{8ex}
   	\subfigure[]{\label{fig:sandwich:2}
    		\centering
    		\includegraphics[width=.36\textwidth]{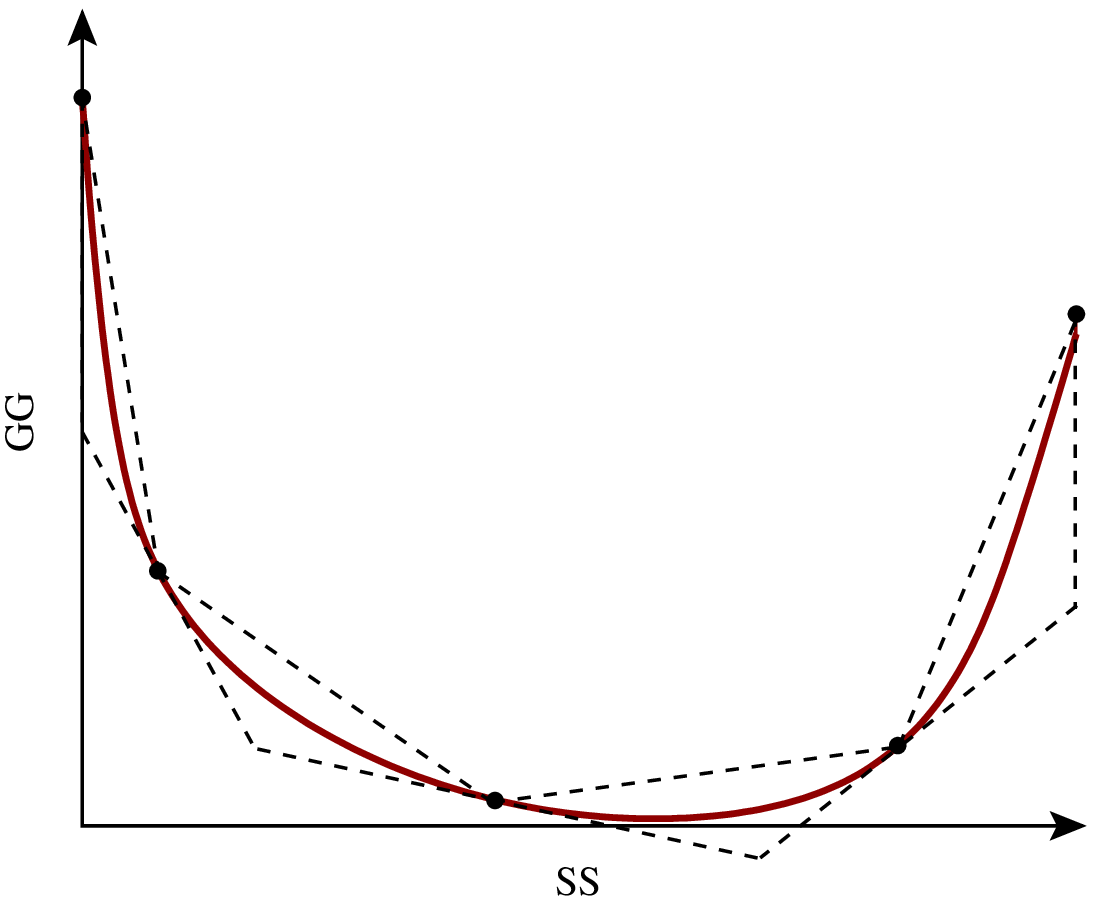}
   	}
    \caption{An illustration of the \emph{sandwich} algorithm. 
            (a) A single refinement step on a segment of the uncertainty curve. 
            (b) Two refinement steps on the full curve.}
\end{figure}

Since $\gu(s)$ is the lower boundary of the convex region $\DD$, it is itself a convex function. We can therefore use the \emph{sandwich} 
algorithm described in \cite{rote_convergence_1992} to approximate it. The algorithm can be easily understood by studying
\fref{sandwich:1}: consider a segment of the curve $\gu(s)$ with two end points $A$ and $B$, whose coordinates are denoted
by $(a, \gu(a))$ and $(b, \gu(b))$, respectively. Also given are \label{page:supporting_line}supporting lines%
 \footnote{A supporting
line is a line that intersects a curve but does not separate any two points on the curve \cite{boyd_convex_2004}.}
containing the end points, represented by the line segments $AD$ and $BD$. Due to the convexity of $\gu(s)$, the chord that connects $A$ to $B$ must lie entirely above the curve and thus form an upper bound. Similarly, the combination of $AD$ and $BD$ forms a piecewise linear lower bound of $\gu(s)$. 

To refine these two initial bounds, let $\alpha$ be the slope of the chord, \emph{i.e.},
\begin{equation}\label{eq:alpha_chord}
\alpha = \frac{\gu(b) - \gu(a)}{b - a}.
\end{equation}
Computing the smallest eigenvalue $q(\alpha)$ and the associated eigenvectors of $\mM(\alpha)$, we can obtain a new point on the curve, denoted by $C$ in \fref{sandwich:1}. The $s$-$g$ coordinates of $C$ are $(\vv^T \mL \vv, \vv^T \mP_{u_0}^2 \vv)$, where $\vv$ is a unit-norm element in the eigenspace $\Ss(\alpha)$. Our previous analysis in \sref{gamma_s}---in particular, \eref{halfplane} and \eref{eigen_v}---guarantees that the line
\[
g - \alpha s = q(\alpha),
\]
which passes through $C$, must be a supporting line of $\gu(s)$.
In other words, $\alpha$ is a subderivative of $\gu(s)$ at point $C$, and is the derivative if it exists. This property, together with the construction of $\alpha$ in \eref{alpha_chord},
also ensures that $C$ is always located between $A$ and $B$.
As illustrated in the figure, the curve is now bounded above by joining the three points ($A$, $C$ and $B$), and it is bounded below by joining the three supporting lines ($AE, EF$ and $FB$).


The above procedure can then be repeated, in a recursive fashion, on the two curve segments $AC$ and $CB$.  Each stage of the recursion roughly doubles the number of points in the approximation, and we proceed until a fixed number of refinements have been computed. \fref{sandwich:2} shows the lower and upper bounds of $\gu(s)$ obtained by starting from two initial points $(0, \vf_1^T \mP_{u_0}^2 \vf_1)$ and $(\lambda_{N}, \vf_N^T \mP_{u_0}^2 \vf_N)$ and running the algorithm for two refinement iterations, involving a total of five eigenvalue evaluations (each corresponding
to a single point drawn on the curve.) We can see that the proposed algorithm starts producing reasonable approximations of $\gu(s)$ after just a small number of steps.

Let $\eta_u^{(n)}(\cdot)$ and $\eta_\ell^{(n)}(\cdot)$ denote, respectively, the upper and lower bounds the algorithm generates after $n$ eigenvalue evaluations. We measure the quality of approximation by computing the Hausdorff distance \cite{rote_convergence_1992} between these two bounds, defined as
\begin{equation*}
    d(n) = \sup_{s_1} \underset{s_2}{\inf\vphantom{\sup}}
    \left[(s_1 - s_2)^2 + (\eta_u^{(n)}(s_1) - \eta_\ell^{(n)}(s_2))^2\right]^\frac{1}{2}.
\end{equation*}
Informally, the Hausdorff distance $d(n)$ is small if the two bounding curves are close to each other. 
The following theorem, which follows directly from \cite[Theorem 3]{rote_convergence_1992}, shows that $d(n)$ is of order $1/n^2$.

\begin{theorem}\label{thm:sandwich}
    Let $\varepsilon > 0$ be any preset precision level. To get $d(n) \le \varepsilon$, it is sufficient to run the approximation algorithm until we have $n \ge \max\set{4, \sqrt{9 W/ \varepsilon}+2}$, where $W = \sqrt{\lambda_{N}^2 + \mathcal{E}_G^4(u_0)}$.
\end{theorem}

\begin{remark}
    In many practical applications, the underlying graph $G$ is large but sparse. Correspondingly, $\mM(\cdot)$ are sparse matrices. Obtaining an approximation of $\gu(s)$ within a given precision $\varepsilon$ then boils down to computing (\emph{e.g.}, via iterative power methods) the smallest eigenvalue and an associated eigenvector of about $\mathcal{O}(1/\sqrt{\varepsilon})$ sparse matrices. 
\end{remark}

Instead of approximating the whole curve, we may wish to find $\gu(s)$ only 
for some particular value of $s$, as well as the signal that achieves it. 
The sandwich algorithm can be modified slightly to this end.  At each step of 
the approximation procedure, we can choose to refine 
only the segment containing $s$, ignoring all other segments. Iterating in this way, 
we will find both $\gu(s)$ and the vector with spectral spread $s$ that achieves the bound.


\section{The Uncertainty Curve for Special Graph Families}
\label{sec:special}

The uncertainty curves for several standard graph families are analyzed in this section. The structure and regularity of complete graphs and 
star graphs make it possible to find closed-form expressions for their corresponding curves. For \Erdos-\Renyi{} random graphs
\cite{Erdos:1959ai,Erdos:1960dq}, we will derive and compute analytical approximations for the expected (\emph{i.e.}, mean)
curves under different parameters. Throughout this section, the distance metric $d(\cdot, \cdot)$ is assumed to be the geodesic distance.

\subsection{Complete Graphs}

A complete graph is a fully-connected graph in which every pair of distinct vertices is connected by an edge \cite{Kolaczyk:2009}. 
It is often used to model fully-connected subgraphs, or \emph{cliques}, in real-world networks \cite{newman_networks:_2010}.
The Laplacian matrix of a complete graph with $N$ vertices is given by
\begin{equation}
\mL_{ij} = \begin{cases}
1, &\text{if } i = j;\\
-\frac{1}{N-1}, &\text{otherwise},
\end{cases}
\label{eq:complete_laplacian}
\end{equation}
\emph{i.e.}, the diagonal of $\mL$ is all $1$, and the off-diagonal elements are all equal to $-\frac{1}{N-1}$.
It is easy to verify that $\mL$ has eigenvalue $0$ with multiplicity $1$, and eigenvalue $\frac{N}{N-1}$ with multiplicity $N-1$.
Without loss of generality, we can choose the first vertex as the center. The diagonal distance matrix is then
\begin{equation}
    \mP_{u_0} = \diag\{0, 1, 1, \ldots, 1\}. \label{eq:complete_dist_matrix}
\end{equation}

We would like to compute the uncertainty curve $\gamma(s)$ for a complete graph for $s \in [0, \frac{N}{N-1}]$.
First, we will show that any vector that achieves the uncertainty curve has a special form.
\begin{proposition}\label{prop:constant_nodes}
For a complete graph, suppose $\widetilde{\vx}$ achieves the uncertainty curve. Then $\widetilde{\vx}$ is of the form
\begin{equation}\label{eq:eigenvector_constant}
\widetilde{\vx} = [x_1, x_2, x_2, \ldots, x_2]^T.
\end{equation}
\end{proposition}
\begin{IEEEproof}
See Appendix~\ref{appendix:constant_nodes}.
\end{IEEEproof}

The result in Proposition~\ref{prop:constant_nodes} suggests that, for complete graphs, we need only consider vectors of the form in \eref{eigenvector_constant}. Enforcing the unit-norm constraint on \eref{eigenvector_constant}, we can further simplify these eigenvectors as $\widetilde{\vx}(\theta) = [\cos \theta, \frac{\sin \theta}{\sqrt{N-1}},\frac{\sin \theta}{\sqrt{N-1}},\ldots,\frac{\sin \theta}{\sqrt{N-1}}]^T$ for some parameter $\theta$. 
The graph spread in this case is given by
\[
\DG = \sum_{i=1}^{N-1} 1 \cdot \frac{\sin^2 \theta}{N-1} = \frac{1}{2} - \frac{1}{2} \cos 2\theta,
\]
where the second equality is due to a standard trigonometric identity. Meanwhile, by using the variational form in \eref{norm_var}, we can compute the spectral spread as
\begin{align}
    \DS & = (N-1) \left(\frac{\cos\theta}{\sqrt{N-1}} - \frac{\sin\theta}{N-1}\right)^2\nonumber\\
           & = \frac{N}{2N-2} - \frac{1}{\sqrt{N-1}} \sin 2 \theta + \frac{N-2}{2N-2} \cos 2\theta.     \label{eq:complete_spec}
\end{align}

Combining these two expressions and using the identity $\sin^2 2\theta + \cos^2  2\theta = 1$, we can see that the uncertainty curve $\gu(s)$ is part of the ellipse given by
\begin{equation}
    (2\DG - 1)^2 + (N-1)\left(\DS + \frac{N-2}{N-1} \DG - 1\right)^2 = 1. \label{eq:complete_ellipse}
\end{equation}
For fixed $s = \DS$, solving for $\gu(s) = \DG$ [by picking the smaller of the two solutions to (\ref{eq:complete_ellipse})] leads to
\begin{align}
&\gu(s) = \nonumber \\
&  \frac{N - s(N-2) - 2 \sqrt{1 - (N-2)(s-1) - (N-1)(s-1)^2}}{4 + (N-2)^2/(N-1)},
\label{eq:complete_curve}
\end{align}
for $0 \le s \le \frac{N}{N-1}$. Thus, the curve is the entire lower half of the ellipse given by (\ref{eq:complete_ellipse}).
When the graph is large (\emph{i.e.}, $N \gg 1$), 
this curve converges to a straight line $\gu(s) = 1 - s$ in the $s$--$g$ plane.

\subsection{Star Graphs}

A star graph \cite{GodsilR:2001} with $N$ vertices has one central vertex and $N-1$ leaves, each connected by a single edge to the center. It is a prototypical example
of a hub in a network \cite{newman_networks:_2010}. The Laplacian matrix can be expressed in block form as
\begin{equation}
    \mL = \begin{pmatrix}
        1                       &    -\frac{1}{\sqrt{N-1}} \vone_{N-1}^T \\
        -\frac{1}{\sqrt{N-1}}\vone_{N-1}   &   \mI_{N-1} 
    \end{pmatrix},
    \label{eq:star_laplacian}
\end{equation}
where $\vone_{N-1}$ is the $(N-1)$-vector of all ones, and $\mI_{N-1}$ is the $(N-1) \times (N-1)$ identity matrix.
Since the graph is bipartite, the largest eigenvalue of $\mL$ is always equal to 2 \cite{chung_spectral_1997}.
Let $u_0$ be the center of the star; the diagonal distance matrix is again given by $\mP_{u_0} = \diag\{0, 1, 1, \ldots, 1\}$.

Just as for the complete graph, we can always represent signals that achieve the uncertainty curve on star graphs 
as $\widetilde{\vx}(\theta) = [\cos \theta, \frac{\sin \theta}{\sqrt{N-1}},\frac{\sin \theta}{\sqrt{N-1}},\ldots,\frac{\sin \theta}{\sqrt{N-1}}]^T$
for some $\theta$ (see the remark in Appendix \ref{appendix:constant_nodes} for justification). Now, the graph spread is
given by $\DG = \sin^2 \theta = \frac{1}{2} - \frac{1}{2} \cos 2 \theta$; again, by using \eref{norm_var},
the spectral spread can be computed as
\begin{align*}
    \DS & = (N-1) \left(\frac{\cos\theta}{\sqrt{N-1}} - \frac{\sin\theta}{\sqrt{N-1}}\right)^2\\
        & = 1 - \sin 2 \theta.
\end{align*}
The lower bound curve is thus the lower part of the ellipse defined by 
\[
    \left(\DS - 1\right)^2 + (2\DG - 1)^2 = 1.
\]
Written explicitly, the curve is
\begin{equation}
\gu(s) = \frac{1}{2} \left(1 - \sqrt{s(2-s)}\right), \quad \text{for } 0 \le s \le 2.
\label{eq:star_curve}
\end{equation}
We note that, unlike the complete graph case, this curve does not depend on the size of the graph.

\subsection{\Erdos-\Renyi{} Random Graphs}
An \Erdos-\Renyi{} random graph $G$ is generated by taking $N$ vertices and selecting each pair of vertices
to be an edge with probability $p$, independent of all other potential edges.
We denote by $\GG_p(N,p)$ the statistical ensemble of the resulting graphs. First studied by \Erdos{} and \Renyi{} \cite{Erdos:1959ai,Erdos:1960dq}, $\GG_p(N,p)$ may be the simplest random graph model. Although they do not capture all of the behaviors of real networks, \Erdos-\Renyi{} graphs are an excellent theoretical model because they lend themselves
to tractable analysis.

\begin{figure*}[t]
    \begin{center}
        \input{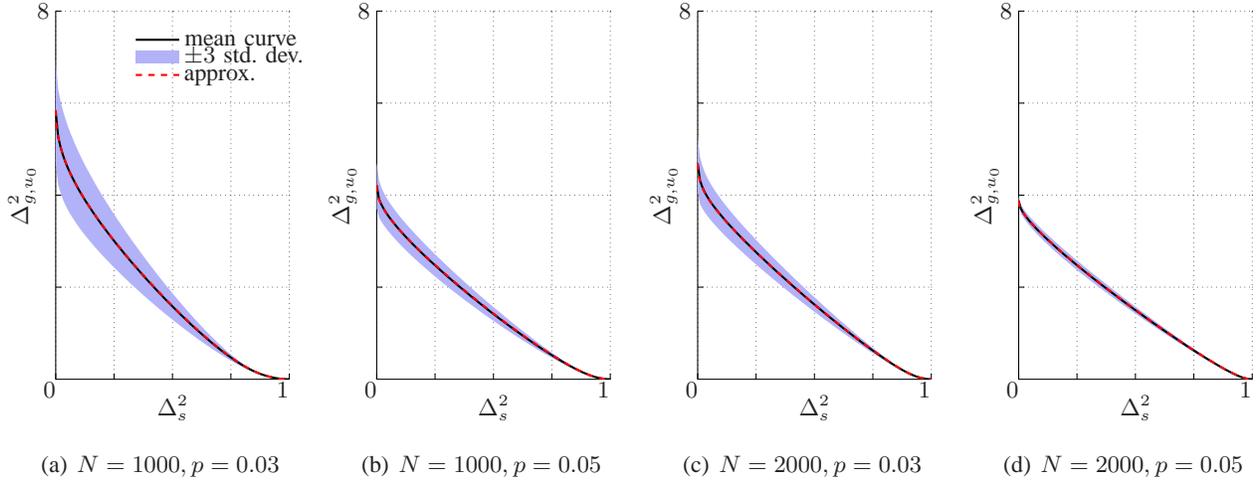}
    \end{center}
    \caption{Uncertainty curves for \Erdos-\Renyi{} graphs. For each choice of $(N,p)$ parameters, 1000 \Erdos-\Renyi{} graphs
    were generated and their uncertainty curves for $s \in [0,1]$ were computed using the sandwich approximation procedure described in Section \ref{sec:Bound}. 
    The geodesic distance function is used.
    Each curve was interpolated to generate comparable curves on a regular grid. For each parameter choice, the mean and standard deviation of the interpolated curve was computed 
    over the ensemble. The mean curve is plotted on the graphs as a solid line, with shaded areas illustrating the three
    standard deviation levels. Meanwhile, the approximate expected value computed before generating the curves is plotted as a dashed red line.
    The shape of the uncertainty curve is clearly quite stable across each ensemble, especially as $N$ and $p$ increase,
    and the approximate expectation curve is quite accurate.}
    \label{fig:ER_curve}
\end{figure*}

To study the properties of the uncertainty curves for \Erdos-\Renyi{} graphs, we generated several realizations drawn from $\GG_p(N,p)$ and used the approximation algorithm described in Section \ref{sec:curve_approx} 
to compute their uncertainty curves. It quickly emerged that the curves for different realizations generated with the same parameters were, for reasonable sizes of $N$, tightly clustered around a common mean curve. This is illustrated in Figure \ref{fig:ER_curve}, which shows the mean curves and estimated standard deviations for several parameter values. In what follows, we develop an analytic approximation for computing the expected (\emph{i.e.} mean) uncertainty curve for different choices of parameters $N$ and $p$.

Recall from the definition of the uncertainty curve that we are trying to approximate the expectation of
\begin{equation}\label{eq:gamma_random}
\gu(s) = \min_{\vx \in \ell^2(G)} \vx^T \mP_{u_0}^2 \vx \quad\text{subject to } \norm{\vx}^2 = 1 \text{ and } \vx^T \mL \vx = s
\end{equation}
over random graphs drawn from $\GG_p(N,p)$. The matrices $\mP_{u_0}^2$ and $\mL$ and the optimal vector $\vx$ that solves the minimization problem are all random quantities. Since $\gu(s)$ is obtained through a nonconvex quadratic program, there is generally no closed-form expressions linking $\gu(s)$ to $\mP_{u_0}^2$ and $\mL$. As a result, directly computing the expectation of $\gu(s)$ will be difficult. To make the problem tractable, we proceed by replacing $\vx^T \mP_{u_0}^2 \vx$ and $\vx^T \mL \vx$ in \eref{gamma_random} with their respective expected values and minimizing after the fact. Later we will see that this strategy turns out to be very effective in generating accurate approximations.

Another observation that emerged from our numerical experiment was a characteristic of the vectors that achieved the bound with $s \leq 1$: these vectors were all approximately radial functions, \emph{i.e.}, the value at any vertex $v$ was a function of $d(u_0, v)$.
Because this simplifies the analysis greatly, we will only consider the part of the curve with $s \leq 1$,
which corresponds to signals that are maximally localized in both the graph and spectral domains.
We will explicitly incorporate this assumption by focusing on vectors whose values depend only on distance
from $u_0$. In this case, the original $N$-dimensional vector $\vx \in \ell^2(G)$ can be represented by a smaller vector $\vy$, with $x(v) = y(d(u_0,v))$. The dimensionality of $\vy$ is equal to $\mathcal{E}_G(u_0) + 1$, where $\mathcal{E}_G(u_0)$ is the eccentricity of the center vertex. We note that $\mathcal{E}_G(u_0)$ is a random variable that in principle can take any value between $0$ and $N-1$. When $N$ is large, however, we can find a small number $d_{\max} \sim \mathcal{O}(\log N)$ such that $\mathcal{E}_G(u_0) \le d_{\max}$ with high probability \cite{albert_statistical_2002}. So, in what follows, we will treat $\vy$ as a vector in $\RR^{d_{\max}+1}$.


For a given, deterministic $\vy$, we will compute the expectations (over the randomness of the graph model) of $\norm{\vx}^2$ and $\vx^T \mP_{u_0}^2 \vx$. To that end, we define $f_d$ as the probability that 
a vertex $v$ chosen uniformly at random from $V\backslash\{u_0\}$ has a distance $d(u_0,v) = d$. 
The special case $f_1 = p$ is  easy to verify. For the other cases, we will use the results of Blondel
\textit{et al.} \cite{blondel_distance_2007}, who developed a recursive formula%
\footnote{Unlike our construction, they allowed $v$ to be any vertex in $V$, including $u_0$; thus, in their result, 
$f_0 = \frac{1}{N}$, and all other values of $f_d$ differ from ours by a factor of $\frac{N-1}{N}$. For large $N$
the difference is negligible.} 
to find (approximate) analytical expressions of the entire sequence $\set{f_d}$.
The expected number of vertices at a distance $d \geq 1$ is $(N-1) f_d$. It follows that, for fixed $\vy$,
\begin{equation}\label{eq:approx_norm}
\mathbb{E}\left[\norm{\vx}^2\right] = \mathbb{E}\left[\sum_{v \in V} y(d(u_0,v))^2\right] \approx y^2(0) + \sum_{k=1}^{d_{\max}} (N-1) f_k y^2(k)
\end{equation}
and
\begin{equation}\label{eq:approx_graph_spread}
\mathbb{E}\left[\vx^T \mP_{u_0} \vx\right] = \mathbb{E}\left[\sum_{v \in V} d(u_0,v)^2 x(v)^2\right] \approx \sum_{k=1}^{d_{\max}} k^2  (N-1) f_k y^2(k),
\end{equation}
where the approximations are due to the truncation of $\vy$ at dimension $d_{\max}$.

The spectral spread is more complicated. We start with the expression
\[
\vx^T \mL \vx = \sum_{u \sim v} \left(\frac{x(u)}{\sqrt{\deg(u)}} - \frac{x(v)}{\sqrt{\deg(v)}} \right)^2.
\]
By assuming that the degree of every vertex is approximately equal to its expectation $(N-1)p$, we write
\begin{equation}
\vx^T \mL \vx \approx \frac{1}{(N-1) p} \sum_{u \sim v} (x(u) - x(v))^2.
\label{eq:er_ds1}
\end{equation}
Recall that $x(v) = y(d(u_0,v))$. Consequently, the only edges that contribute to (\ref{eq:er_ds1}) are those between vertices at \emph{different} distances from $u_0$. Since a vertex at distance $k$ can only be connected to vertices at a distance of $k-1$ and $k+1$, we simply need to characterize $M_{k,k+1}$, the expected number of edges from vertices at a distance $k$ to vertices at a distance $k+1$, for $k = 0$ to $d_{\max}-1$. The expected value of the spectral spread can then be obtained as
\begin{equation}
\mathbb{E}\left[\vx^T \mL \vx\right] \approx \frac{1}{(N-1)p}\sum_{k=0}^{d_{\max}-1} M_{k,k+1} \big(y(k+1) - y(k)\big)^2.
\label{eq:approx_spec_spread}
\end{equation}

It is easy to see that $M_{0,1} = (N-1)p$, since that is simply the expected number of edges incident upon $u_0$. The other terms of $M_{k, k+1}$ can be approximated through a recurrence relation. First, we observe that the expected number of vertices at distance $k$ is $(N-1) f_k$ and the expected number of vertices \emph{not} at distance $k$ (not counting $u_0$) is 
$(N-1) (1-f_k)$. Thus, we can approximate the total number of \emph{potential} edges between these two 
disjoint sets of vertices is $(N-1)^2 f_k(1-f_k)$. Since each potential edge will be chosen with probability
$p$, we get that $M_{k-1,k} + M_{k,k+1} \approx (N-1)^2 p f_k (1-f_k)$, which leads to the following approximate
recurrence relation
\begin{equation}
\begin{cases}
M_{0,1} = (N-1)p \\
M_{k,k+1} \approx (N-1)^2 p f_k (1-f_k) - M_{k-1,k}, & \text{for } k \geq 1.
\end{cases}
\end{equation}

The expressions in \eref{approx_norm}, \eref{approx_graph_spread}, and \eref{approx_spec_spread} show that the expected values of the squared norm, graph spread, and spectral spread are all nonnegative quadratic forms involving the vector $\vy \in \RR^{d_{\max} + 1}$. It follows that we can write
\[
\mathbb{E}\left[\norm{\vx}^2\right] \approx \vy^T \mH_a \vy, \ \ \mathbb{E}\left[\vx^T \mP_{u_0} \vx\right] 
\approx \vy^T \mP^2_a \vy, \text{ and } \mathbb{E}\left[\vx^T \mP_{u_0} \vx\right]  \approx \vy^T \mL_a \vy,
\]
for some positive semidefinite matrices $\mH_a, \mP^2_a, \mL_a$, respectively. Substituting these expectations for their (random) counterparts in \eref{gamma_random}, we compute our approximation of the expected uncertainty curve, $\tgu(s)$, as
\begin{equation}\label{eq:gamma_expectation}
    \tgu(s) = \min_{\vy \in \RR^{d_{\max}+1}} \vy^T \mP_a^2 \vy \quad\text{subject to } \vy^T \mH_a \vy = 1 \text{ and } \vy^T \mL_a \vy = s.
\end{equation}
We note that this minimization problem (a quadratic program with quadratic constraints) has exactly the same mathematical structure as the one previously studied in \eref{qcqp}. Using the same techniques derived in \sref{gamma_s}, we can show that any solution to \eref{gamma_expectation} satisfies the (generalized) eigenvalue problem
\begin{equation}\label{eq:gen_eig}
(\mP_a^2 - \alpha \mL_a)\vy = \tau_{\min}(\alpha) \mH_a \vy
\end{equation}
for some value of $\alpha$, where $\tau_{\min}(\alpha)$ is the smallest (generalized) eigenvalue. As before, we can construct a sandwich approximation to the curve by solving \eref{gen_eig} for a sequence of $\alpha$'s.


Despite the various approximations made along the way, the analytical solution obtained in \eref{gamma_expectation} 
fits experiment remarkably well. As illustrated in Figure \ref{fig:ER_curve}, the resulting analytic curves 
(shown in dashed lines) match almost perfectly with the observed sample average (shown in solid lines). 
We note that the matrices in \eref{gamma_expectation} are of size $d_{\max} \times d_{\max}$, which is much 
smaller than $N \times N$. For example, for the $\GG_p(10^6, 10^{-4})$ model, we would have $d_{\max} = 4$ (the
smallest $d$ such that $1 - \sum_{k=1}^{d} f_k < 10^{-7}$.)

Thus, the analytic approximation derived here can be computed far faster than the actual uncertainty curve 
for any realization of the model, and does not itself require any realization to be generated.

\section{Diffusion Processes and Uncertainty Bounds}
\label{sec:diffusion}

In constructing dictionaries to represent signals on graphs, one would like the dictionary elements to be localized in both graph and spectral domains. Quantifying the signal localization in these two domains and studying their fundamental tradeoff have been one of the motivations of this work. To test the theoretical results and the computational algorithm presented in \sref{Bound}, we consider two graph wavelet transforms in the literature: the diffusion wavelets of Coifman and Maggioni \cite{coifman_diffusion_2006} and the spectral graph wavelet transform of Hammond \emph{et al.} \cite{hammond_wavelets_2010}. The localization properties of these two constructions are studied on a graph visualized in \fref{boundcurve:1}
based on the network of football games played in the 2000 regular season by NCAA Division I-A teams \cite{girvan_community_2002}. While the spectral graph wavelet transform does not downsample the graph, the diffusion wavelet transform does. 
In our experiment, the center vertex $u_0$ is chosen to be one of the vertices that remain in the downsampled graph at the coarsest level of the diffusion wavelet transform.

\begin{figure}[t]
	\centering
	\psfrag{DELS2}{\footnotesize $\Delta_s^2$}
	\psfrag{DELG2}[c][t]{\footnotesize $\Delta_{g,u_0}^2$}
    \psfrag{U0}{\footnotesize $u_0$}
    \psfrag{10}[r][r]{\footnotesize $10$}
    \psfrag{5}[r][r]{\footnotesize $5$}
    \psfrag{0}[rb][rb]{\footnotesize $0$}
    \psfrag{00}[lt][lt]{\footnotesize $0$}
    \psfrag{1}[l][l]{\footnotesize $1$}
	\subfigure[]{\label{fig:boundcurve:1}
		\centering
		\includegraphics[width=0.36\textwidth]{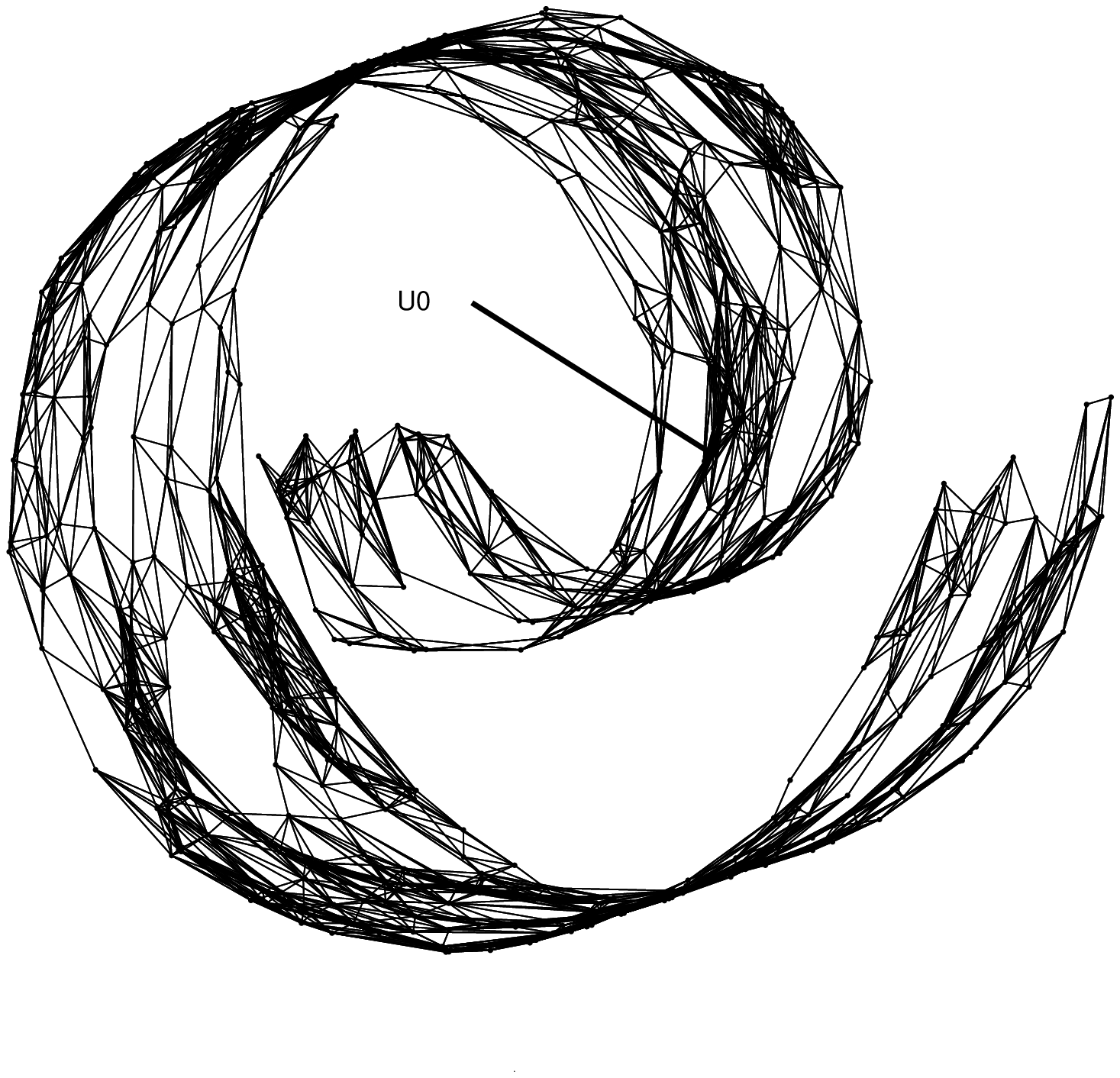}%
	}
	\hspace{6ex}
	\subfigure[]{\label{fig:boundcurve:2}
		\centering
		\includegraphics[width=.42\textwidth]{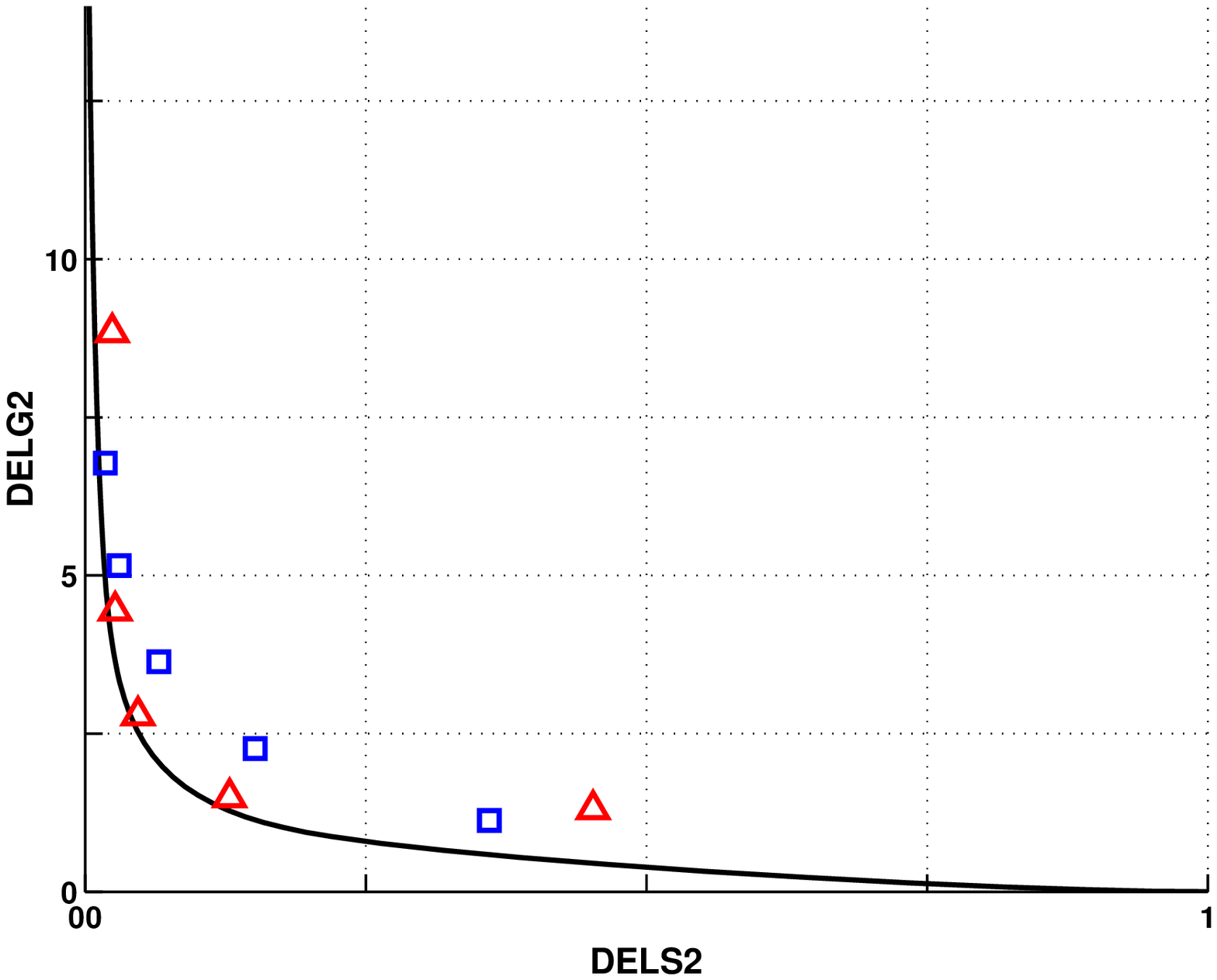}
	}
    \caption{(a) Network of football games between NCAA Division I-A teams in the 2000 regular season \cite{girvan_community_2002}; (b) Spectral spread versus graph spread on this graph.
    Solid line: computed uncertainty curve $\gu(s)$. Triangles: scaling functions in diffusion 
    wavelets \cite{coifman_diffusion_2006}. Squares: scaling functions
    in spectral graph wavelet transform (SGWT) \cite{hammond_wavelets_2010}. 
    (The true SGWT scaling functions are not related to the wavelet functions by a two-scale relation; here,
    we simply take the cumulative sum of the coarsest-level scaling function and higher-level wavelet functions.)
}
\end{figure}

\fref{boundcurve:2} shows several scaling functions from both constructions plotted against the uncertainty curve $\gu(s)$, with the latter obtained by using the sandwich algorithm in \sref{curve_approx}. In this and all subsequent experiments, we use eight refinement iterations (for a total of \label{page:niters} $257$ sparse eigenvalue evaluations)
to plot the uncertainty curves. At this level, we find the lower and upper approximations of $\gu(s)$ to be visually indistinguishable.
As predicted, both the spectral graph wavelet and diffusion wavelet constructions result in basis elements that obey the computed bound. In fact, they follow the curve quite well.

The diffusion wavelets are based on the evolution of a discrete time
diffusion process on a graph. In the classical setting, where the signal domain is the real line,
there is a strong connection between the \emph{continuous time} diffusion
process and the Heisenberg uncertainty curve: to see this, consider a diffusion (\emph{i.e.} heat) equation
\begin{equation}\label{eq:diffusion_1d}
\frac{\partial u}{\partial t} = \frac{\partial^2 u}{\partial y^2},
\end{equation}
where $u(y,t)$ is a function of $y,t \in \RR$. This equation governs the conduction of heat in physical processes, and its solution was the original motivation for Fourier analysis. The fundamental solution to \eref{diffusion_1d}, \emph{i.e.}, the solution with the initial condition that $u(y,0) = \delta(y-y_0)$ for a given $y_0$, is the Gaussian kernel
\[
K(t,y,y_0) = \frac{1}{\sqrt{4 \pi t}} e^{-\frac{(y-y_0)^2}{4 t}}.
\]
Thus, if we start with an impulse and evolve according to \eref{diffusion_1d}, at time $t$ we get a function with
time spread $t$ and frequency spread $\frac{1}{4t}$, achieving the classical Heisenberg uncertainty $\Delta_t^2 \, \Delta_\omega^2 \geq \frac{1}{4}$ with equality. In other words, the diffusion kernels on the real line are exactly the signals that achieve the time-frequency uncertainty bound.

This line of thought motivated us to consider a continuous-time diffusion process on graphs, governed by an equation analogous to (\ref{eq:diffusion_1d}):
\begin{equation}\label{eq:diffusion_graph}
    \frac{d \vx}{dt} = - \mL \vx,
\end{equation}
where $\mL$ is the graph Laplacian. With the initial condition $\vx(0) = \vimp{u_0}$, the solution to \eref{diffusion_graph} is \cite{coifman_diffusion_2006b}
\begin{equation}
    \vx(t) = e^{-t \mL} \vimp{u_0} = \sum_{i=1}^N e^{-t \lambda_i} \vf_i \vf_i^T \vimp{u_0},
    \label{eq:diffusion_solution}
\end{equation}
where $e^{-t \mL}$ is the matrix exponential of $\mL$, $\set{\lambda_i}$ are the eigenvalues of $\mL$, and $\set{\vf_i}$ are the corresponding eigenvectors.
Denote by $\etau(s)$ the curve in the $s$--$g$ plane traced out by the diffusion process. The curve can be given in parametric form as
\begin{equation}
    \begin{cases}
        s(t) = \frac{\vx(t)^T \mL \vx(t)}{||\vx(t)||^2} \\
        \etau(s) = \frac{\vx(t)^T \mP_{u_0}^2 \vx(t)}{||\vx(t)||^2}.
        \label{eq:diffcurve}
    \end{cases}
\end{equation}
We show in Appendix \ref{appendix:diffusion2} that $s(t)$
is a strictly decreasing function of $t$; therefore it is one-to-one. Furthermore, $s(0) = 1$ and $\lim_{t \to \infty} s(t) = 0$.
All together, this guarantees that the function $\etau(s)$ is well-defined for every $s \in (0,1]$.


\begin{figure*}[t]
    \input{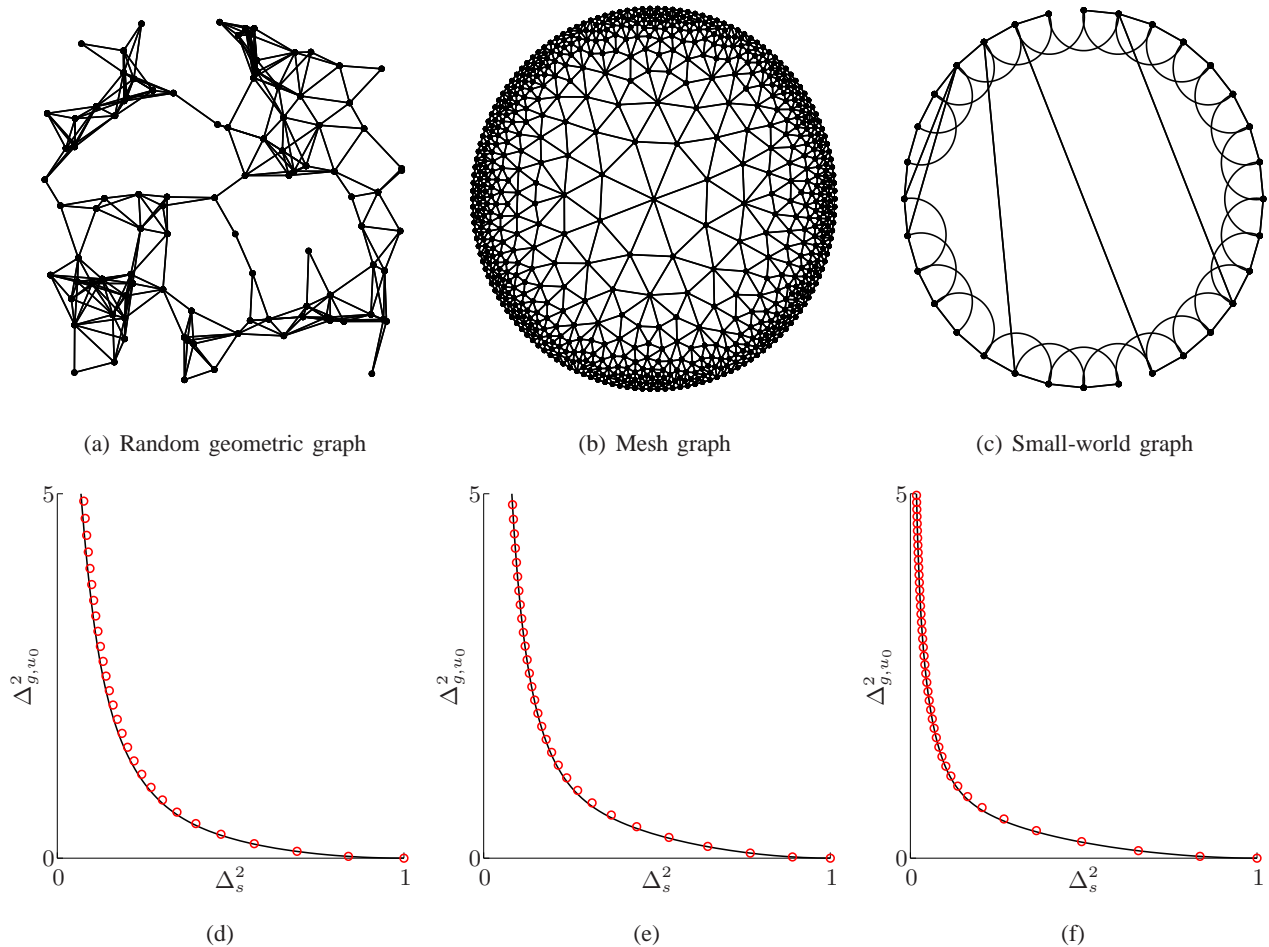}
    \caption{Diffusion process versus the uncertainty curve for three types of graph. (a) A
    random geometric graph \cite{penrose_random_2003}, (b) a triangular mesh \cite{strang_analysis_2008}, and (c) a small-world graph \cite{watts_collective_1998}. Below
    each graph, (d), (e), and (f) show the associated uncertainty curves (solid black line). A continuous-time diffusion process
    is run on each graph, beginning with an impulse at one vertex, and the resulting spreads are plotted (solid red line with circles). 
    The circles are evenly spaced in time. The
    diffusion process tracks the curve closely, though close examination reveals that the match is not exact.}
    \label{fig:diffusion}
\end{figure*}

We plot in \fref{diffusion} the diffusion curve $\etau(s)$ and the uncertainty curve $\gu(s)$
for three different graphs: a random geometric graph \cite{penrose_random_2003} that can capture the 
connectivity of wireless sensor networks; an unstructured triangular mesh%
\footnote{This graph was generated using the Mesh2D MATLAB toolbox written by Darren Engwirda, available online
    at MATLAB Central (\url{http://www.mathworks.com/matlabcentral/fileexchange/25555}).}
for finite element analysis \cite{strang_analysis_2008}; and a small-world graph \cite{watts_collective_1998} that 
serves as the mathematical model for social and various other empirical networks. The geodesic distance function is used.
In all three cases, the spreads of the diffusion process, though not exactly achieving the bounds as in the 
classical setting, match the uncertainty curves remarkably well.


The following proposition, proved in Appendix \ref{appendix:diffusion}, asserts that for certain special graphs the
match between $\etau(s)$ and $\gu(s)$ is exact.
\begin{proposition}
    \label{prop:diffusion}
     For all $s \in (0, 1]$, $\etau(s) = \gu(s)$ if (a) $G$ is a complete graph with $N$ vertices and $u_0$ is any vertex; or (b) $G$ is a star
    graph with $N$ vertices and $u_0$ is the vertex with degree $N-1$.
\end{proposition}

For general graphs we can show that, under certain conditions, the low-order derivatives of the uncertainty curve and the diffusion curve match.
\begin{proposition}
    \label{prop:diffusion2}
    Let $G$ be any connected graph and $u_0$ be any vertex on $G$. Then $\etau(1) = \gu(1) = 0$,
    $\left.\frac{d\etau}{ds}\right|_{s=1} = \left.\frac{d\gu}{ds}\right|_{s=1} = 0$,
    and 
    \begin{align}
    \left.\frac{d^2\gu}{ds^2}\right|_{s=1} & = \frac{\deg{u_0}}{2 \sum_{v \sim u_0} \frac{1}{d(v,u_0)^2 \deg{v}}} 
    \leq \left.\frac{d^2\etau}{ds^2}\right|_{s=1} \nonumber \\ &= 
    \frac{\deg{u_0}}{2} \frac{\sum_{v \sim u_0} \frac{d(v, u_0)^2}{\deg{v}}}{\left(\sum_{v \sim u_0} \frac{1}{\deg{v}} \right)^2},
    \label{eq:2nd_deriv_matches}
    \end{align}
    with equality \emph{if and only if} $d(v,u_0)$ is identical for every $v \sim u_0$.
\end{proposition}

This proposition is proved in Appendix \ref{appendix:diffusion2}. It is easy to verify that the geodesic distance
satisfies the condition required for equality in (\ref{eq:2nd_deriv_matches}).
Extrapolating the observations in Figure \ref{fig:diffusion} and results in Propositions \ref{prop:diffusion} and \ref{prop:diffusion2}
leads us to believe that diffusion kernels on arbitrary graphs will always be close to optimal in graph 
and spectral localizations.
We leave further rigorous study of this tantalizing conjecture as an important line of future work.

\section{Conclusions}
\label{sec:conclusions}

Analogous to the classical Heisenberg uncertainty principle in time-frequency analysis, an uncertainty principle for signals defined on graphs 
was developed in this work. After presenting quantitative definitions of the signal ``spreads'' in the graph and spectral domains,
we provided a complete characterization of the feasibility region achieved by these two quantities. The lower boundary of the region,
which is analogous to the classical uncertainty bound (\ref{eq:classical_uncertainty}), was shown to be achieved by eigenvectors associated 
with the smallest eigenvalues of a particular matrix-valued function. Furthermore, the convexity of the uncertainty curve allows it to be 
efficiently approximated by solving a sequence of eigenvalue problems. We derived closed-form formulas of the uncertainty curves 
for complete graphs and star graphs, and developed a fast analytical approximation for the expected uncertainty curve for 
\Erdos-\Renyi{} random graphs. The localization properties of two existing wavelet transforms were evaluated. 
Finally, numerical experiments and analytical results led us to an intriguing connection between diffusion processes on graphs and the uncertainty bounds.

\appendix

\subsection{The convexity of $\DD$}
\label{appendix:convexity}
We would like to prove that the set $\DD$ is convex as long as the number of vertices $N \geq 3$. (The need for such a condition will be made clear shortly.) This is equivalent to showing the following result.

\begin{proposition}
    \label{prop:defconvex}
    Suppose that there exist two vectors $\vx_1, \vx_2$ in $\RR^N$ with $N \geq 3$, such that
    \begin{equation}\label{eq:affine}
       \vx_i^T \vx_i = 1, \quad \vx_i^T \mL \vx_i = s_i,  \text{ and } \vx_i^T \mP_{u_0}^2 \vx_i = g_i, \qquad \text{for } i = 1, 2. \\
    \end{equation}
Then for any $\beta \in [0,1]$, we can always find a vector $\vx$ in $\RR^N$ satisfying
\begin{equation}\label{eq:convexity}
        \vx^T \vx = 1,  
        \quad   \vx^T \mL \vx = s, 
        \text{ and }\vx^T \mP_{u_0}^2 \vx = g,
\end{equation}
where $s \bydef \beta s_1 + (1-\beta) s_2$ and $g \bydef \beta g_1 + (1-\beta) g_2$.
\end{proposition}

We will prove the above proposition by recasting the problem in $\op{Sym}_N$, the Hilbert space of real, symmetric
$N \times N$ matrices. The space is endowed with the Hilbert-Schmidt inner product defined by
$\langle \mA, \mB \rangle_{\text{HS}} \bydef \op{tr}(\mA^T \mB) = \op{tr}(\mA \mB)$, where $\text{tr}(\cdot)$ is the trace of a matrix. Every $\vx \in \RR^N$ can be mapped onto a matrix $\mX = \vx \vx^T$ in $\op{Sym}_N$. Finding a vector $\vx$ satisfying the conditions in \eref{convexity} then boils down to finding a rank-one positive semidefinite matrix $\mX = \vx \vx^T$ satisfying the following three constraints
\begin{equation}\label{eq:convexity_m}
        \op{tr}(\mX)= 1, \quad \op{tr}(\mL \mX) = s \quad \text{and }
        \op{tr}(\mP_{u_0}^2 \mX) = g.
\end{equation}

The requirement that $\mX$ be a rank-one matrix makes this a hard problem, because the cone of rank-one matrices
is not convex. Instead, we will use the following theorem to relax the problem to the cone of positive semidefinite matrices $\Ss^N_+$, which is convex.
\begin{theorem}[Barvinok \cite{barvinok_remark_2001}]\label{thm:barvinok}
    Suppose that $R > 0$ and $N \geq R + 2$. Let $\HH \subset \op{Sym}_N$ be an affine subspace such that $\op{codim}(\HH) \leq \dbinom{R+2}{2}$.
    If the intersection $\Ss^N_+ \cap \HH$ is nonempty and bounded, then there is a matrix $\mX$ in $\Ss^N_+ \cap \HH$ such that $\op{rank}(\mX) \leq R$.
\end{theorem}

\begin{IEEEproof}[Proof of Proposition~\ref{prop:defconvex}]
First, we note that the three equalities in \eref{convexity_m} are all affine constraints on $\mX$. Together, they define a hyperplane $\HH \subset \op{Sym}_N$ with $\op{codim}(\HH) \leq 3 = \dbinom{1 + 2}{2}$. (In fact, $\mI$, $\mL$, and $\mP_{u_0}^2$ are linearly independent, so $\op{codim}(\HH) = 3$.) To apply Theorem~\ref{thm:barvinok}, we verify next that $\Ss^N_+ \cap \HH$ is nonempty and bounded.

First we show that it is bounded: let $\mX$ be an arbitrary matrix in the intersection $\Ss^N_+ \cap \HH$ (assuming one exists),
and let $\set{\nu_1, \nu_2, \ldots, \nu_N}$ be its eigenvalues.
The equalities $1 = \op{tr}(\mX) = \sum_{n=1}^N \nu_n$, together with the nonnegativity of the eigenvalues, imply that
\[
\norm{\mX}_{\text{HS}}^2 = \op{tr}(\mX^2) = \sum_{n=1}^N \nu_n^2 \le \sum_{n=1}^N \nu_n = 1.
\]
Therefore, $\Ss^N_+ \cap \HH$ is a subset of the unit ball in $\op{Sym}_N$ and is thus bounded.

To show that $\Ss^N_+ \cap \HH$ is nonempty, we explicitly construct a member of the set. Let $\vx_1, \vx_2$ be the two vectors 
satisfying \eref{affine}. On mapping the vectors to two matrices $\mX_1 \bydef \vx_1 \vx_1^T$ and  $\mX_2 \bydef \vx_2 \vx_2^T$,
the constraints in \eref{affine} can be rewritten as 
\[
\op{tr}(\mX_i) = 1, \quad \op{tr}(\mL \mX_i) = s_i \quad \text{and }\op{tr}(\mP_{u_0}^2 \mX_i) = g_i, \qquad \text{for } i = 1, 2.
\]
$\mX_1$ and $\mX_2$ are both in $\Ss^N_+$. Now set $\mX' = \beta \mX_1 + (1 - \beta) \mX_2$. It is easy to see that $\mX' \in \HH$ and, because $\Ss^N_+$ is convex, $\mX' \in \Ss^N_+$ as well. To be sure, the matrix $\mX' \in \Ss^N_+ \cap \HH$ is not necessarily of rank one. However, the result of Theorem~\ref{thm:barvinok} (for the case when $R = 1$) guarantees the existence of a rank one matrix $\mX$ in $\Ss^N_+ \cap \HH$. Decomposing this matrix as $\mX = \vx \vx^T$ and using the equivalence between \eref{convexity} and \eref{convexity_m}, we can conclude that the resulting vector $\vx$ satisfies all the constraints in \eref{convexity}.
\end{IEEEproof}

\begin{remark}
The above proof uses Theorem~\ref{thm:barvinok} for the case when $R = 1$.  Consequently, we need to work with $N \ge R + 2 = 3$. This requirement is sharp in that the achievable region $\DD$ for a graph with two vertices (\emph{i.e.}, $N=2$) is not convex. The only connected graph with $N=2$ is the complete graph. All unit-norm signals on this graph can be parametrized as $(\cos \theta, \sin \theta)$. By computing the corresponding graph Laplacian and distance matrices, it is easy to show that the achievable region is only the boundary of an ellipse (not including its interior) and hence is not convex.
\end{remark}


\subsection{Proof of Lemma~\ref{lem:hprops}}
\label{appendix:diff_analysis}

(a) For any $\alpha_1 < \alpha_2$, let $\vv_1$ and $\vv_2$ be two unit-norm eigenvectors in $\Ss(\alpha_1)$ and $\Ss(\alpha_2)$, respectively.
Applying Rayleigh's inequality, we get $\vv_2^T \mM(\alpha_1) \vv_2 \ge q(\alpha_1) = \vv_1^T \mM(\alpha_1) \vv_1$
Similarly, we have $-\vv_2^T \mM(\alpha_2) \vv_2 \geq - \vv_1^T \mM(\alpha_2) \vv_1$. A combination of these two inequalities leads to
\begin{equation}\label{eq:h_decreasing}
\vv_2^T \left(\mM(\alpha_1) - \mM(\alpha_2)\right) \vv_2 \ge \vv_1^T \left(\mM(\alpha_1) - \mM(\alpha_2)\right) \vv_1.
\end{equation}
Recall that $\mM(\alpha) = \mP_{u_0}^2 - \alpha \mL$, and therefore $\mM(\alpha_1) - \mM(\alpha_2) = (\alpha_2 - \alpha_1) \mL$. 
Replacing this identity into \eref{h_decreasing}, we thus have
\[
\vv_2^T  \mL \vv_2 \ge \vv_1^T  \mL \vv_1.
\]
Note that $\vv_1$ and $\vv_2$ can be arbitrary unit-norm elements in $\Ss(\alpha_1)$ and $\Ss(\alpha_2)$, respectively. If, in particular, 
we choose $\vv_1, \vv_2$ to be those that attain the maximization in \eref{hpm}, we get $h_{+}(\alpha_2) = \vv_2^T  \mL \vv_2 \ge \vv_1^T  \mL \vv_1 = h_{+}(\alpha_1)$. Similarly, we can show that $h_{-}(\alpha_2) \ge h_{-}(\alpha_1)$. 

(b) We will only consider the limits when $\alpha$ tends to $-\infty$ as given in \eref{limit_right}. The other case, when $\alpha$ tends
to $+\infty$, can be analyzed in a similar way, and its proof will be omitted. Let $\alpha > 0$ be any positive number. By definition,
\begin{equation}\label{eq:hp_lb}
h_{+}(\alpha) \ge h_{-}(\alpha) \ge 0,
\end{equation}
where the second inequality is due to the Laplacian matrix $\mL$ being positive semidefinite. Next, we show that $h_{+}(\alpha)$ 
can be made arbitrarily close to $0$ as $\alpha \rightarrow -\infty$. To that end, let $\vv$ be any unit-norm eigenvector in $\Ss(\alpha)$, and $\vf_1$ be the first eigenvector of $\mL$ as defined in \eref{dc}. Since $\Ss(\alpha)$ is associated with the smallest eigenvalue $q(\alpha)$, we have, from Rayleigh's inequality,
\[
\vv^T (\mP_{u_0}^2 - \alpha \mL) \vv \le \vf_1^T (\mP_{u_0}^2 - \alpha \mL) \vf_1 = \vf_1^T \mP_{u_0}^2 \vf_1,
\]
with the equality coming from the identity $\mL \vf_1 = 0$. For any $\alpha < 0$, rearranging the above expression leads to
\begin{equation}\label{eq:ds_bound}
    \vv^T \mL \vv \le -\frac{1}{\alpha} \left(\vf_1^T \mP_{u_0}^2 \vf_1 - \vv^T \mP_{u_0}^2 \vv\right) \le -\frac{\mathcal{E}_G^2(u_0)}{\alpha},
\end{equation}
where the second inequality uses the bound of the graph spread as provided in Proposition~\ref{prop:feasible_props}. 
Since \eref{ds_bound} holds for \emph{any} nonzero element $\vv$ from $\Ss(\alpha)$, we must have $h_{+}(\alpha) \le -\mathcal{E}_G^2(u_0) / \alpha$, which, when combined with \eref{hp_lb}, completes the proof.

(c) First, using eigenvalue perturbation results, we will derive a candidate set $\calA$ of points such
that $q(\alpha)$ is certainly analytic on $[a,b] \backslash \calA$. We will show that $\calA$ is finite, so that the set 
of nonanalytic points of $q(\alpha)$ is finite as well. Then, we will compute $h_-(\alpha)$ and $h_+(\alpha)$ explicitly, and show
that they are are left- and right-continuous, respectively, and that they are equal to the negative left- and right-derivatives
of $q(\alpha)$, respectively.
We will then show that $h_-(\alpha) = h_+(\alpha)$ everywhere except a subset $\calB \subseteq \calA$; therefore, they satisfy (\ref{eq:hpm_derivative}).
Since $\calA$ is finite, it follows that $\calB$ is finite as well.

The starting point of our analysis is the following result.
\begin{proposition}
    \label{prop:eig_functions}
There exist $N$ analytic functions $\lambda_1(\cdot), \ldots, \lambda_N(\cdot)$ and $N$
analytic vector-valued functions $\vx_1(\cdot), \ldots, \vx_N(\cdot)$ such that
\begin{equation}\label{eq:eigen_equation}
\mM(\alpha) \vx_i(\alpha) = \lambda_i(\alpha) \vx_i(\alpha),
\end{equation}
and $\vx_i(\alpha)^T \vx_j(\alpha) = \delta_{ij}$.
\label{prop:analyticfuncs}
\end{proposition}
\begin{IEEEproof}
Standard perturbation results \cite[p. 404]{lancaster_theory_1985} guarantee the existence of such functions for any matrix function that is analytic and whose value is always Hermitian. The function $M(\cdot)$ as defined in \eref{M_alpha} is affine in $\alpha$, and thus analytic; it is symmetric and real for every $\alpha$, and thus Hermitian. Therefore functions with the
properties listed in the proposition do exist.
\end{IEEEproof}

From Proposition \ref{prop:eig_functions}, we can write $q(\alpha)$ as
\begin{equation}\label{eq:q_min}
                q(\alpha) = \min_{1 \leq i \leq N} \lambda_i(\alpha),
\end{equation}
where the $\set{\lambda_i(\cdot)}_i$ are the eigenvalue functions guaranteed by the proposition. For any $\alpha_0 \in \RR$, if $S(\alpha_0)$ has dimension one, then precisely one of the eigenvalue functions is equal to $q(\cdot)$ at $\alpha_0$, say $\lambda_k(\alpha_0) = q(\alpha_0)$. Pick some $\varepsilon < \frac{1}{2} \min_{j \neq k} |\lambda_j(\alpha_0) - \lambda_k(\alpha_0)|$. Since every $\lambda_j(\cdot)$ is analytic, we can find some neighborhood $\mathcal{N}$ of $\alpha_0$ for which $|\lambda_j(\alpha) - \lambda_j(\alpha_0)| < \varepsilon$ for every $j$. This guarantees that $\lambda_k(\alpha) < \lambda_j(\alpha)$ on $\mathcal{N}$ for every $j \neq k$. Thus $q(\alpha) = \lambda_k(\alpha)$ on $\mathcal{N}$.
Since $\lambda_k(\cdot)$ is analytic on $\mathcal{N}$, we have that $q(\cdot)$ is analytic on $\mathcal{N}$ and therefore at $\alpha_0$.  We can make this more general. Suppose instead of only one eigenvalue function attaining the minimum at $\alpha_0$, there are multiple eigenvalue functions [\emph{e.g.}, two, denoted by $\lambda_{k_1}(\cdot)$ and $\lambda_{k_2}(\cdot)$] that attain the minimum, and that they are all equal on a neighborhood $\mathcal{N}$ of $\alpha_0$. All the other eigenvalue functions are larger at $\alpha_0$. Again, the analyticity allows us to find
            a neighborhood $\mathcal{N}' \subseteq \mathcal{N}$ on which all the other eigenvalue functions are larger than $\lambda_{k_1}(\cdot) = \lambda_{k_2}(\cdot)$.
            Now, since $q(\alpha) = \lambda_{k_1}(\alpha) = \lambda_{k_2}(\alpha)$, the function $q(\alpha)$ is analytic on $\mathcal{N}'$ as well.

Thus, a necessary condition for $q(\cdot)$ to be nonanalytic at $\alpha_0$ is that two (or more) \textit{distinct} eigenvalue functions must intersect at $\alpha_0$. 
Define $\mu_j(\cdot), j = 1, \ldots, N', N' \leq N$ as the set of distinct eigenvalue functions, and let $n_j$ be the multiplicity of the eigenvalue function $\mu_j(\cdot)$.
Now consider an arbitrary finite interval $[a, b]$ and define
\begin{equation*}
    \calA = \bigcup_{1 \leq i < j \leq N'} \set{\alpha \in [a, b]: \mu_i(\alpha) = \mu_j(\alpha)}.
\end{equation*}
It is a well known property of analytic functions that if they are equal on more than a finite set of points in an 
interval, then they are identical.  Since the $\mu_i(\cdot)$ are distinct analytic functions, no two of them can 
be equal on more than a finite set of points in $[a, b]$. Thus $\calA$ is the finite union of finite sets, and therefore contains
only a finite number of points

Next, we connect $q(\alpha)$ to $h_{+}(\alpha)$ and $h_{-}(\alpha)$. At any point $\alpha_0 \in [a, b]$, there can be 
$k \ge 1$ distinct eigenvalue functions that achieve the minimum in \eref{q_min}. Without loss of generality, we shall assume they are the first $k$ functions, 
$\mu_1(\cdot), \ldots, \mu_k(\cdot)$. The associated eigenvectors, $\vx_{ij}(\alpha_0)$, for $i = 1, \ldots, k$ and $j = 1, \ldots, n_i$,
form an orthonormal basis for the eigenspace $\Ss(\alpha_0)$. Any unit-norm element $\vv \in \Ss(\alpha_0)$ can
then be written as $\vv = \sum_{i=1}^k \sum_{j=1}^{n_i} c_{ij} \, \vx_{ij}(\alpha_0)$, for some constant coefficients $\set{c_{ij}}$ satisfying $\sum_{i=1}^k \sum_{j=1}^{n_i} c_{ij}^2 = 1$.

We now define an analytic function $\vv(\alpha) \bydef \sum_{i=1}^k \sum_{j=1}^{n_i} c_{ij} \, \vx_{ij}(\alpha)$, with $\vv(\alpha_0) = \vv$. 
The eigenvalue identity in \eref{eigen_equation} implies that
$\mM(\alpha) \vv(\alpha) = \sum_{i=1}^k \sum_{j=1}^{n_i} c_{ij} \mu_i(\alpha) \vx_{ij}(\alpha)$. Differentiating both sides of this equality yields
\begin{equation}\label{eq:eigen_equation_diff}
    \mM'(\alpha) \, \vv(\alpha) + \mM(\alpha) \vv'(\alpha) = \sum_{i=1}^k\sum_{j=1}^{n_i} c_{ij} \mu'_i(\alpha) \,\vx_{ij}(\alpha) + \sum_{i=1}^k\sum_{j=1}^{n_i} c_{ij} \mu_i(\alpha) \vx'_{ij}(\alpha).
\end{equation} 
Evaluating \eref{eigen_equation_diff} at $\alpha = \alpha_0$, pre-multiplying it by $\vv^T(\alpha_0)$ and using the substitutions
$\mM'(\alpha) = -\mL$, $\mM(\alpha_0) \vv(\alpha_0) = q(\alpha_0) \vv(\alpha_0)$, and $\mu_i(\alpha_0) = q(\alpha_0)$ for every $i$, we get
\begin{align}
-\vv^T(\alpha_0) \mL \vv(\alpha_0) + q(\alpha_0) \vv^T(\alpha_0) \vv'(\alpha_0) 
& = \sum_{i=1}^k \sum_{j=1}^{n_i} c_{ij}^2 \mu'_i(\alpha_0) + q(\alpha_0) \sum_{i=1}^k \sum_{j=1}^{n_i} c_{ij} \vv^T(\alpha_0) \vx_{ij}'(\alpha_0).
    \label{eq:eigen_equation_diff_alpha0}
\end{align}
The second terms on the left-hand and right-hand sides of \eref{eigen_equation_diff_alpha0} are equal, leaving us with
\begin{equation}\label{eq:ss_diff}
\vv^T(\alpha_0) \mL \vv(\alpha_0)
= -\sum_{i=1}^k\sum_{j=1}^{n_i} c_{ij}^2 \mu'_i(\alpha_0).
\end{equation}
By definition, $h_{+}(\alpha_0)$ and $h_{-}(\alpha_0)$ are the two extreme values of $\vv^T(\alpha_0) \mL \vv(\alpha_0)$.
Maximizing (and minimizing) the quantity in \eref{ss_diff} subject to the unit-norm constraint $\sum_{i=1}^k\sum_{j=1}^{n_i} c_{ij}^2 = 1$, we have
\begin{equation}\label{eq:hpm_diff}
h_+(\alpha_0) = \max_{1\le i \le k} (-\mu'_i(\alpha_0)) \quad \text{and} \quad h_{-}(\alpha_0) = \min_{1\le i \le k} (-\mu'_i(\alpha_0)).
\end{equation}
Now, there must exist some $m,\ell \in \set{1, \ldots, k}$ such that
\begin{equation}\label{eq:q_ml}
q(\alpha) = \begin{cases}
	     \mu_\ell(\alpha) & \text{if } \alpha \leq \alpha_0 \\
         \mu_m(\alpha) & \text{if }\alpha \geq \alpha_0
\end{cases}
\end{equation}
on some neighborhood $\mathcal{N}$ of $\alpha_0$, which can be chosen to be small enough that $\calN \cap \calA = \set{\alpha_0}$ if $\alpha_0 \in \calA$ or
$\calN \cap \calA = \emptyset$ otherwise. We must have $\mu'_m(\alpha_0) = \min_{1\leq i \leq k } \mu'_i(\alpha_0)$,
since if $\mu'_{j}(\alpha_0) < \mu'_m(\alpha_0)$ for some $j$, then on a sufficiently small neighborhood of $\alpha_0$,
we would have $q(\alpha) = \mu_{j}(\alpha) < \mu_m(\alpha)$ for $\alpha > \alpha_0$, contradicting (\ref{eq:q_ml}).%
\footnote{The requirement that  $\mu'_m(\alpha_0) = \min_{1\leq i \leq k } \mu'_i(\alpha_0)$ might not always be sufficient to uniquely determine $m$, however.
In the case that multiple distinct eigenvalue functions achieve the minimum derivative,
$\mu_m(\cdot)$ is then determined by comparing the higher order derivatives. This nuance does not affect our proof, which only depends on the first derivative.}
Meanwhile, away from $\alpha_0$ there are no other points in $\calN$ at which multiple distinct eigenvalue functions intersect.
Thus, from (\ref{eq:hpm_diff}), we have that $h_+(\alpha) = -\mu'_m(\alpha)$ on $\calN \cap [\alpha_0, \infty)$. Since the $\mu_i(\cdot)$ are
all analytic, $h_+(\alpha)$ is right-continuous at $\alpha_0$. Furthermore, since $q(\alpha) = \mu_m(\alpha)$ on $\calN \cap [\alpha_0, \infty)$,
$h_+(\alpha_0)$ is equal to the negative right-derivative of $q(\alpha)$ at $\alpha_0$. By similar arguments, we can show that
$h_-(\alpha)$ is left-continuous at $\alpha_0$
and is equal to the negative left-derivative of $q(\alpha)$ at $\alpha_0$.

A necessary condition for $h_-(\alpha_0) \neq h_+(\alpha_0)$ is that $k > 1$, \emph{i.e.}, there are multiple distinct eigenvalue
functions achieving the minimum in (\ref{eq:q_min}). Thus the set of points $\calB$ at which they differ satisfies $\calB \subseteq \calA$, so
$\calB$ is finite. Meanwhile, if $h_+(\alpha_0) = h_-(\alpha_0)$, then the equality must hold for all $\alpha \in \calN$ as well because of the way we constructed the neighborhood $\calN$.
Since $h_-(\alpha)$ is left-continuous and $h_+(\alpha)$ is right-continuous at $\alpha_0$, both functions are continuous at $\alpha_0$. Equality also means
the left- and right-derivatives of $q(\alpha)$ are equal at $\alpha_0$, and thus
$q'(\alpha_0)$ is well-defined with $h_+(\alpha_0) = h_-(\alpha_0) = -q'(\alpha_0)$.

\subsection{Proof of Proposition~\ref{prop:constant_nodes}}
\label{appendix:constant_nodes}

For $N=2$ the proposition is trivial, so let us assume $N > 2$. 
By Theorem \ref{thm:smallesteig}, $\widetilde{\vx}$ must be an eigenvector associated with the smallest
eigenvalue of $\mM(\alpha) = \mP_{u_0}^2 - \alpha \mL$ for some $\alpha$, where $\mL$ and $\mP_{u_0}$
are given by (\ref{eq:complete_laplacian}) and (\ref{eq:complete_dist_matrix}), respectively.
$\mM(\alpha)$ is given in block form as
\begin{equation*}
   \mM(\alpha) = \left[ \begin{array}{c|c}
        -\alpha & \frac{\alpha}{N-1}\vec{1}_{N-1}^T \\
    \hline
        \frac{\alpha}{N-1}\vec{1}_{N-1} & \mB
    \end{array}\right],
\end{equation*}
where $\mB$ is the $(N-1) \times (N-1)$ circulant matrix 
$\mB = \left(1 - \frac{N}{N-1}\alpha \right) \mI_{N-1} + \frac{\alpha}{N-1} \vec{1}^{\vphantom{T}}_{N-1} \vec{1}_{N-1}^T$.
Let $\set{\vw_1, \ldots, \vw_{N-2}}$ be an orthonormal set of vectors in $\RR^{N-1}$ such that $\vw_i \perp \vec{1}_{N-1}$. This
set spans the subspace of vectors in $\RR^{N-1}$ orthogonal to $\vec{1}_{N-1}$. 
It is easy to verify that $\mB \vw_i = (1-\frac{N}{N-1}\alpha)\vw_i$. Furthermore, if we set $\vv_i = (0, \vw_i^T)^T$,
then we can see that $\vv_i$ are all eigenvectors of $\mM(\alpha)$ with eigenvalue $1 - \frac{N}{N-1}\alpha$.

If we can show that this is not the smallest eigenvalue of $\mM(\alpha)$, \textit{i.e.}~that $q(\alpha) \neq 1 - \frac{N}{N-1}\alpha$,
then it follows that $\widetilde{\vx}$ [an eigenvector of $\mM(\alpha)$ corresponding to $q(\alpha)$] must be orthogonal
to every $\vv_i$ for $i=1, \ldots, N-2$. This will then guarantee that $\widetilde{\vx}$ is of the form (\ref{eq:eigenvector_constant}).


To show that $q(\alpha) \neq 1 - \frac{N}{N-1}\alpha$, we let $\vy = \left[y_1, \ldots, y_N\right]$ be chosen
such that $||\vy|| = 1$, $y_1 \neq 0$, and $\vy^T \vec{1}_N = 0$. This last property makes $\vy$ an eigenvector
of $\mL$ with eigenvalue $\frac{N}{N-1}$. We have $\vy^T \mP_{u_0}^2 \vy = 1 - y_1^2$
and $\vy^T \mL \vy = \frac{N}{N-1}$. Thus $\vy^T \mM(\alpha) \vy = 1 - y_1^2 - \frac{N}{N-1}\alpha < 1 - \frac{N}{N-1}\alpha$.
It follows from the Rayleigh inequality that $q(\alpha) \leq \vy^T \mM(\alpha) \vy < 1 - \frac{N}{N-1}\alpha$,
proving the proposition.

%

\textit{Remark}: With small modifications, the above proof can be used to demonstrate that the same property holds
for star graphs, \textit{i.e.}~any vector achieving the uncertainty curve must be of the form in (\ref{eq:eigenvector_constant}).
For a star graph with $N$ vertices, we have
\begin{equation}
\mM(\alpha) =    \left[ \begin{array}{c|c}
    -\alpha & \frac{\alpha}{\sqrt{N-1}}\vec{1}_{N-1}^T \\
    \hline
    \frac{\alpha}{\sqrt{N-1}}\vec{1}_{N-1} & (1-\alpha) \mI_{N-1}
    \end{array}\right].
\end{equation}
Again, there is an $(N-2)$--dimensional eigenspace spanned by the same set $\set{\vv_i}_{i=1}^{N-2}$ as in
the complete graph case above. In this case, the eigenvalue associated with that subspace is $1-\alpha$.
Thus, to show that the smallest eigenvector is of the desired form, we must simply show that there is
some unit norm vector $\vy$ for which $\vy^T \mM(\alpha) \vy < 1 - \alpha$, guaranteeing that the eigenvector
associated with the smallest eigenvalue is orthogonal to the eigenspace spanned by $\set{\vv_i}_{i=1}^{N-1}$.
Our test vector here is $\vy = (1, 0, \ldots, 0)^T$,
which gives us $\vy^T \mM(\alpha) \vy = -\alpha < 1- \alpha$, so the same property holds for the star graph
as the complete graph.

\subsection{Proof of Proposition \ref{prop:diffusion}}
\label{appendix:diffusion}
(a) Let $\set{\vf_1, \ldots, \vf_N}$ be an orthonormal basis of $\RR^N$ with $\vf_1 = \frac{1}{\sqrt{N}} \vec{1}_N$.
It is easy to verify that these are eigenvectors of $\mL$ [given in (\ref{eq:complete_laplacian})] 
with corresponding eigenvalues $\lambda_1 = 0$ and $\lambda_k = \frac{N}{N-1}$
for $k = 2, \ldots, N$.

It follows from (\ref{eq:diffusion_solution}) that the diffusion process starting from $\vx_0 = \vimp{u_0}$ can be obtained as
\begin{align}
    \vx(t) & = \vf_1 \vf_1^T \vimp{u_0} + e^{-\lambda_2 t} (\mI - \vf_1 \vf_1^T) \vimp{u_0}.
    \label{eq:complete_process_solution}
\end{align}
Assuming without loss of generality that $u_0 = 1$ and using the fact that $\vf_1 = \frac{1}{\sqrt{N}} \vec{1}$, we have 
$\vx(t) = \frac{1}{N} \left[1 + (N-1)e^{-\lambda_2 t}, 1 - e^{-\lambda_2 t}, \ldots, 1 - e^{-\lambda_2 t} \right]^T$.
Using our knowledge of $\mL$ and the fact that $\mP_{u_0} = \diag(0, 1, \ldots, 1)$,
it is now straightforward to compute the spreads as
$\DS(\vx(t))  = \dfrac{N e^{-2 \lambda_2 t}}{1 + (N-1)e^{-2 \lambda_2 t} }$ and
$\DG(\vx(t))  = \dfrac{\frac{N-1}{N}\left(1 - e^{-\lambda_2 t} \right)^2 }{1 + (N-1)e^{-2 \lambda_2 t}}.$
We can verify that these spreads satisfy (\ref{eq:complete_curve}). Thus, for all $t \geq 0$, $\vx(t)$ achieves the
uncertainty curve. $\DS(\vx(t))$ is continuous and $\lim_{t \to \infty} \DS(\vx(t)) = 0$, so $\etau(s) = \gu(s)$ for $s \in (0,1]$.

(b) Here, we assume without loss of generality that the center of the star, \textit{i.e.},~the vertex with degree $N-1$ is $u_0 = 1$.
Again, we explicitly construct an orthonormal eigenbasis for $\mL$, given in this case by (\ref{eq:star_laplacian}). In what follows,
we will assume that $N > 2$; the star graph with $2$ vertices is the same as the complete graph with $2$ vertices, so
the proof from (a) will apply to that case.
Let $\vf_1 = \left[\frac{1}{\sqrt{2}\vphantom{(N-1)}}, \frac{1}{\sqrt{2(N-1)}}, \ldots, \frac{1}{\sqrt{2(N-1)}} \right]^T$, 
$\vf_N = \left[-\frac{1}{\sqrt{2}\vphantom{(N-1)}}, \frac{1}{\sqrt{2(N-1)}}, \ldots, \frac{1}{\sqrt{2(N-1)}} \right]^T$, 
and $\vf_k = \left[0, \vg_{k}^T\right]^T$ for $k = 2, \ldots, N-1$, where $\set{\vg_k}_{k=1}^{N-1}$ is any orthonormal basis for $\RR^{N-1}$
satisfying $\vg_1 = \frac{1}{\sqrt{N-1}} \vec{1}_{N-1}$.
It is easy to verify that $\set{\vf_k}_{k=1}^{N}$ forms an orthonormal basis for $\RR^N$, and that the $\vf_k$ are eigenvectors
of $\mL$ with corresponding eigenvalues $\lambda_1 = 0$, $\lambda_2 = \cdots = \lambda_{N-1} = 1$, and $\lambda_N = 2$.

Similar to (\ref{eq:complete_process_solution}), we can compute the diffusion process explicitly as
\begin{align}
    \vx(t) & = \vf_1 \vf_1^T \vimp{u_0} + e^{-t} (\mI - \vf_1 \vf_1^T - \vf_N \vf_N^T)\vimp{u_0}  + e^{-2 t} \vf_N \vf_N^T \vimp{u_0} \\
    & = \left(1 - e^{-t}\right) \vf_1 \vf_1^T \vimp{u_0} + \left(e^{-2t} - e^{-t}\right) \vf_N \vf_N^T \vimp{u_0} + e^{-t} \vimp{u_0}.
\end{align}
Using the expressions for $\vf_1$ and $\vf_N$, we find that 
$\vx(t) = \left[ \frac{1}{2} + \frac{1}{2}e^{-2t},  \frac{1}{2\sqrt{N-1}} \left( 1 - e^{-2t}\right)\vec{1}_{N-1}^T \right]^T.$ 
From this, we can compute the graph spread as
$\Delta_{g, u_0}^2(\vx(t)) = \dfrac{(1 - e^{-2t} )^2 } {2(1 + e^{-4t})}$
and the spectral spread as
$\DS(\vx(t)) = \dfrac{2 e^{-4 t}}{1 + e^{-4 t}}.$
It is easy to verify that these spreads satisfy (\ref{eq:star_curve}), and so $\vx(t)$ achieves the uncertainty curve for $t \geq 0$.
Once again, $\DS(\vx(t))$ is continuous and $\lim_{t \to \infty} \DS(\vx(t)) = 0$, so $\etau(s) = \gu(s)$ for $s \in (0,1]$.

\subsection{Proof of Proposition \ref{prop:diffusion2}}
\label{appendix:diffusion2}

We know from Theorem \ref{thm:smallesteig} that every point on the uncertainty curve is achieved by an eigenvector
associated with the smallest eigenvalue $q(\alpha)$ of $\mM(\alpha) = \mP_{u_0}^2 - \alpha \mL$. In particular,
the point $(1,0)$ is achieved by $\vimp{u_0}$, which is the eigenvector associated with the matrix $\mM(0) = \mP_{u_0}^2$ 
and eigenvalue $q(0) = 0$.
Since $d(u_0, v) = 0$ if and only if $u_0 = v$ and $d(v,u_0) > 0$ otherwise, the eigenspace $\calS(0)$ is one-dimensional. 
Thus, from the proof of Lemma \ref{lem:hprops} in Appendix \ref{appendix:diff_analysis}, there is some neighborhood $\calN$ of $\alpha = 0$ on which $\calS(\alpha)$ is one-dimensional, and therefore
$q(\alpha)$ is analytic. In this case, there exists some neighborhood of $s = 1$ for which 
we can use the parametric form of the uncertainty curve given in (\ref{eq:gammas_parametric}), namely
$(s, \gu(s)) = (s_u(\alpha), g_u(\alpha))$ where $s_u(\alpha) = -q'(\alpha)$ and $g_u(\alpha) = q(\alpha) - \alpha q'(\alpha)$ for $\alpha \in \calN$.

We can thus compute the derivative of the uncertainty curve parametrically as 
\begin{equation}
\frac{d\gu}{ds}  = \frac{g_u'(\alpha) }{s_u'(\alpha)} = \frac{-\alpha q''(\alpha)}{-q''(\alpha)}  = \alpha \label{eq:dgds}.
\end{equation}
where $\alpha$ is chosen so that $s(\alpha)$ is the argument at which we wish to evaluate the derivative. Similarly, the second derivative is
\begin{align}
    \frac{d^2\gu}{ds^2} = \frac{ \frac{d}{d\alpha}\left(\frac{g_u'(\alpha)}{s_u'(\alpha)}  \right)}{s_u'(\alpha)}
    = \frac{1}{-q''(\alpha)}. \label{eq:d2gds2}
\end{align}

Both (\ref{eq:dgds}) and (\ref{eq:d2gds2}) require that $q''(\alpha)$ be nonzero. 
In what follows, we will explicitly compute $q''(0)$ and show that $q''(\alpha) \neq 0$ for $\alpha \in \calN'$, where $\calN' \subseteq \calN$.
As described in the proof of Lemma \ref{lem:hprops},
there is an analytic eigenvector function $\vv(\alpha)$ defined in a neighborhood of $\alpha = 0$ such that
\begin{equation}
\mM(\alpha) \vv(\alpha) = q(\alpha) \vv(\alpha),
\label{eq:Mvqv}
\end{equation}
with $\vv(0) = \vimp{u_0}$ and $||\vv(\alpha)||^2 = 1$. The spectral spread function is given by $s_u(\alpha) = \vv(\alpha) \mL \vv(\alpha) = -q'(\alpha)$,
where the second equality is due to (\ref{eq:gammas_parametric}).
So we can compute 
\begin{equation}
    q''(\alpha) = -2 \vv(\alpha) \mL \vv'(\alpha). \label{eq:qpp}
\end{equation}
To compute $\vv'(\alpha)$, we differentiate both sides of (\ref{eq:Mvqv}) and after rearranging terms obtain
\begin{equation}
[\mM(\alpha) - q(\alpha)\mI] \vv'(\alpha)  = \mL \vv(\alpha) + q'(\alpha) \vv(\alpha). \label{eq:vderiveq}
\end{equation}
From (\ref{eq:Mvqv}) and the fact that $\calS(\alpha)$ is one-dimensional on $\calN$, $\mM(\alpha) - q(\alpha) \mI$
has a one-dimensional nullspace spanned by $\vv(\alpha)$. Since $0 = \frac{d}{d\alpha}||v(\alpha)||^2 = 2 \vv(\alpha)^T \vv'(\alpha)$, 
when we multiply both
sides of (\ref{eq:vderiveq}) by the Moore-Penrose pseudoinverse of $\mM(\alpha) - q(\alpha) \mI$, we obtain
\begin{equation}
    \vv'(\alpha) = [\mM(\alpha) - q(\alpha) \mI]^+\mL \vv(\alpha), \label{eq:vpp}
\end{equation}
where we have also used the fact that $[\mM - q(\alpha) \mI]^+ \vv(\alpha) = 0$ to simplify the right-hand side of (\ref{eq:vpp}).

Setting $\alpha = 0$ and using the fact that $q(0) = 0$ and $\vv(0) = \vimp{u_0}$, we have $\vv'(0) = \left(\mP_{u_0}^2\right)^+ \mL \vimp{u_0}$.
Substituting this into (\ref{eq:qpp}), we get $q''(0) = -2 \vimp{u_0}^T \mL \left(\mP_{u_0}^2\right)^+\mL\vimp{u_0} 
= -2 \displaystyle\sum_{v \sim u_0} \frac{(\mL \vimp{u_0})_v^2}{d^2(v, u_0)}$, where $(\mL\vimp{u_0})_v$ is the $v$th entry of $\mL \vimp{u_0}$.
From the definition of the graph Laplacian, we have that for every $v \sim u_0$,
$(\mL \vimp{u_0})_v = \displaystyle\frac{-1}{\sqrt{\deg{u_0}}} \frac{1}{\sqrt{\deg{v}}}$.
Thus,
\begin{equation}
    q''(0) = \frac{-2}{\deg{u_0}} \left( \sum_{v \sim u_0} \frac{1}{d(v,u_0)^2 \deg{v}} \right).
\end{equation}
Since the
graph is connected, $q''(0) \neq 0$, and since $q(\alpha)$ is analytic on $\calN$, there exists a neighborhood $\calN' \subseteq \calN$ containing
$0$ on which $q''(\alpha) \neq 0$ as well. Thus our expressions for the first and second derivatives (\ref{eq:dgds}) and (\ref{eq:d2gds2})
are valid at $s = 1$, which corresponds to $\alpha = 0$.
We obtain $\left.\frac{d \gu}{ds}\right|_{s = 1} = 0$ and the expression for $\left.\frac{d^2 \gu}{ds^2}\right|_{s=1}$ given in (\ref{eq:2nd_deriv_matches}).

To compute the derivatives of the curve $\etau(s)$ traced out by the diffusion process $\vx(t)$, we express it parametrically
in terms of $t$, with $(s, \etau(s)) = (s_d(t), g_d(t))$ where 
$s_d(t) = \frac{\vx(t)^T \mL \vx(t)}{||\vx(t)||^2}$ and $g_d(t) = \frac{\vx(t)^T \mP_{u_0}^2 \vx(t)}{||\vx(t)||^2}$.

We first show that $\dot{s}_d(t) < 0$. To simplify the computation of this and other derivatives, we introduce
the function $R_{\mZ}(t) = \frac{\vx(t)^T \mZ \vx(t)}{||\vx(t)||^2}$ for any fixed matrix
$\mZ$. It is easy to verify that since $\dot{\vx}(t) = -\mL \vx(t)$, 
$\dot{R}_{\mZ}(t) = 2R_{\mZ}(t) R_{\mL}(t) - R_{\mL \mZ}(t) - R_{\mZ \mL}(t)$, where the last two terms in the sum are equal if $\mZ$ is symmetric.
Since we have an explicit solution $\vx(t) = e^{-\mL t} \vimp{u_0}$, we can see that $||\vx(t)|| \neq 0$ for all $t$, so that
$R_{\mZ}(t)$ and its derivative is well-defined.

Since $s_d(t) = R_{\mL}(t)$,
we have $\dot{s}_d(t) = 2(s_d(t)^2 - R_{\mL^2}(t)) = 2\left[(\vx(t)^T \mL \vx(t))^2 - \vx(t)^T \mL^2 \vx(t)\right] < 0$ by
the Cauchy-Schwarz inequality. Equality would hold only if $\mL \vx(t)$ were a multiple of $\vx(t)$---\emph{i.e.}, if $\vx(t)$
were an eigenvector. From (\ref{eq:diffusion_solution}) we can see that this could only occur if $\vimp{u_0}$ itself
were an eigenvector, which is impossible for a connected graph. 
We can directly evaluate $s_d(0) = 1$ and $\lim_{t \to \infty} s_d(t) = 0$; combining this with the above result guarantees
that $s_d(t)$ is a one-to-one function with range $(0,1]$. Thus $\etau(s)$ is well-defined on that domain.


Since $g_d(t) = R_{\mP_{u_0}^2}(t)$, we can compute the derivative $\dot{g}_d(0) = g_d(0) s_d(0) - R_{\mL \mP_{u_0}^2}(0) = 0$.
Thus the diffusion curve's derivative at $s=1$ is given by
\begin{equation}
    \left.\frac{d\etau}{ds}\right|_{s = 1} = \frac{\dot{g}_d(t)}{\dot{s}_d(t)} = 0 = \left.\frac{d\gu}{ds}\right|_{s = 1}.
\end{equation}

Meanwhile, we can simplify the second derivative evaluated at $s=1$, obtaining
\begin{equation}
    \left.\frac{d^2\etau}{ds^2}\right|_{s=1} 
    = \frac{\ddot{g}_d(0)}{\dot{s}_d^2(0)} \label{eq:d2etads2}.
\end{equation}
The first derivative of $s_d(t)$ at $t=0$ can be computed as
\begin{align}
    \dot{s}_d(0) & = 2(s_d(0)^2 - R_{\mL^2}(0)) \nonumber \\
                & = 2\left(1 - ||\mL \vimp{u_0} ||^2\right) 
                 = -2\sum_{v \sim u_0} \frac{1}{\deg{u_0} \deg{v}}. \label{eq:sdot0}
\end{align}

The second derivative of $g_d(t)$ is
\begin{align}
    \ddot{g}_d(t)
    & = 2\dot{g}_d(t)s_d(t) + 2g_d(t) \dot{s}_d(t) - 4s_d(t) R_{\mL \mP_{u_0}^2}(t) + 2R_{\mL^2 \mP_{u_0}^2}(t) + 2 R_{\mL \mP_{u_0}^2 \mL}(t) \label{eq:gdotdot}
\end{align}
At $t=0$, the only nonzero term in (\ref{eq:gdotdot}) is the last one: 
\begin{align}
\ddot{g}_d(0)  &= 2 R_{\mL \mP_{u_0}^2 \mL}(0) 
             = 2 \vimp{u_0}^T \mL \mP_{u_0}^2 \mL \vimp{u_0}  \nonumber \\
             &= 2 \sum_{v \sim u_0} \frac{d(v,u_0)^2}{\deg{u_0}\deg{v}}. \label{eq:gddot0}
\end{align}
Now we can combine (\ref{eq:d2etads2}), (\ref{eq:sdot0}), and (\ref{eq:gddot0}) to obtain the expression for
$\left.\frac{d^2\etau}{ds^2}\right|_{s = 1}$ given in (\ref{eq:2nd_deriv_matches}).
By the Cauchy-Schwartz inequality, $\left(\sum_{v \sim u_0} \frac{1}{\deg{v}} \right)^2 \leq \left( \sum_{v \sim u_0} \frac{d(v, u_0)^2}{\deg{v}}\right)
\left(\sum_{v \sim u_0} \frac{1}{d(v, u_0)^2 \deg{v}}\right)$ with equality \emph{if and only if} $d(v, u_0)^2 = c$ for every $v \sim u_0$,
where $c$ is some constant.
Comparing the expressions for the second derivatives of the uncertainty curve and diffusion curve, we can see that
$\left.\frac{d^2\etau}{ds^2}\right|_{s =1} \geq \left.\frac{d^2\gu}{ds^2}\right|_{s=1}$, with equality \emph{if and only if}
$d(v, u_0)$ is identical for every $v \sim u_0$.

\section*{Acknowledgments}
We thank the associate editor and the anonymous referees, especially referee \#1, for their constructive criticism
and valuable comments, which corrected several errors in the original manuscript and greatly improved the presentation of this paper.
\bibliographystyle{IEEEtran}
\bibliography{library,refs}

\end{document}